\documentclass[acmsmall]{acmart}

\AtBeginDocument{%
  \providecommand\BibTeX{{%
    \normalfont B\kern-0.5em{\scshape i\kern-0.25em b}\kern-0.8em\TeX}}}

\setcopyright{cc}
\setcctype{by-nc-nd}
\acmJournal{PACMHCI}
\acmYear{2026} \acmVolume{10} \acmNumber{6} \acmArticle{CSCW079}
\acmMonth{10} \acmDOI{10.1145/3816927}

\acmISBN{978-1-4503-XXXX-X/18/06}

\usepackage{graphicx}
\usepackage{kotex}

\newcommand{\final}[1]{\textcolor{black}{#1}}

\begin{document}

\title[Enabling Sensitive Conversations with Consent Boundaries: Moa, a Platform for Discussing PhD Advising]{Enabling Sensitive Conversations with Consent Boundaries:\\Moa, a Platform for Discussing PhD Advising Relationships}
% \title[Enabling Sensitive Conversations Around Power Dynamics: Moa, a Consentful Platform for Discussing PhD Advising]{Enabling Sensitive Conversations Around Power Dynamics:\\Moa, a Consentful Platform for Discussing PhD Advising}

% \author{Anonymized}

\author{Jane Im}
\authornote{This work was done while the first author was at the University of Michigan.}
\email{im@cispa.de}
\orcid{0000-0002-9614-6535}
\affiliation{%
  \institution{CISPA Helmholtz Center for Information Security}
  % \streetaddress{P.O. Box 1212}
  \country{Germany}
}

\author{Kentaro Toyama}
\email{toyama@umich.edu}
\orcid{0000-0002-9128-2255}
\affiliation{%
  \institution{University of Michigan}
  % \streetaddress{P.O. Box 1212}
  \country{USA}
}



\begin{abstract}

When an individual is harmed by someone in power, such as a workplace manager, it can help to identify allies---people who would offer advice or supportive action. However, ally discovery is fraught because the very people who might be most relevant---e.g., someone who reports to the same manager---might not be sympathetic and could exacerbate the harm. We examine this problem in the context of PhD students navigating advising challenges and present a social media platform called ``Moa'' that brings together a number of features for facilitating ally discovery. Moa's most novel element is an audience selection process that uses what we call \textit{consent boundaries}, which allow users to flexibly define each post or comment's audience based on factors such as common social identity or lived experience, all while preserving mutual anonymity. A 3-week field study with 47 real-world users showed that Moa's features in combination facilitated sensitive conversations about advising, with \final{7 out of 31 posters (22.6\%)} using consent boundaries. We discuss both our overall ``recipe'' for systems for ally discovery and the benefits of a consent-centered approach to design.

\end{abstract}





\received{13 May 2025}
\received[revised]{13 January 2026}
\received[accepted]{17 March 2026}

\maketitle

\section{Introduction}

\begin{figure*}[t!]
    \centering
    \includegraphics[width=0.92\linewidth]{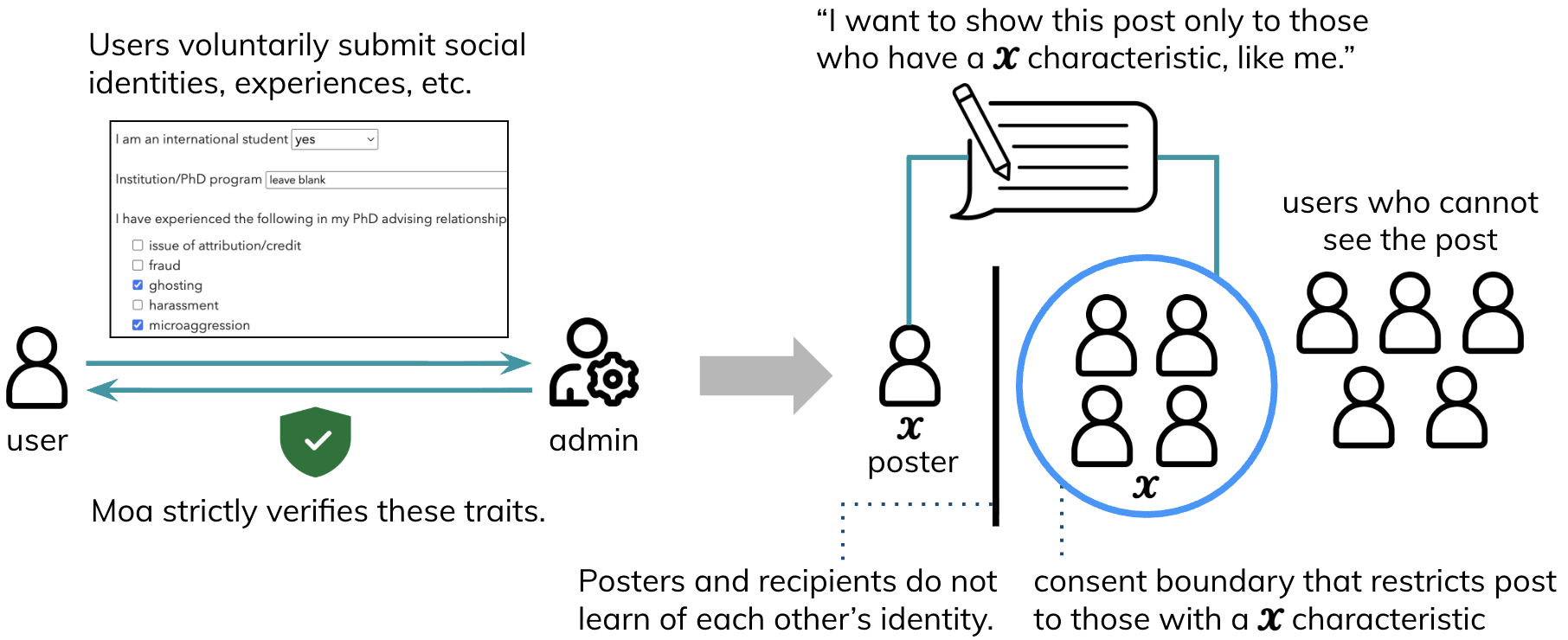}
    \caption{Moa is a social platform that incorporates a novel concept we call a \emph{consent boundary}---it allows users to demarcate precisely who can view their posts in terms of various contextual information, such as social identity, advising challenges, PhD program, and faculty information, while maintaining anonymity{---neither senders nor recipients learn each other's identities}. When a post is created, only users within the specified consent boundary are notified and can view it in their feed.  }
\label{moa-overview}
\end{figure*}

Individuals who suffer from problematic power dynamics, such as mistreatment or abuse from someone in power, can benefit from identifying allies, whether it is to seek advice or remedial action.
Here, we use the word ``ally'' to refer to any individual who is sympathetic and can offer some kind of help, ranging from sharing useful information to actively navigating the situation together.
Example contexts include workplace bullying~\cite{zapf2020empirical}, political persecution~\cite{mcquaid2021surviving}, and sexual harassment~\cite{fox2017metoo}. In contexts where potential allies are initially unknown, however, someone must make at least a partial disclosure of their experience for what we call \emph{ally discovery}\footnote{{We have coined this phrase because literature in political science, collective action, and social movements does not appear to have a phrase representing individuals searching for and finding previously unknown allies. Related terms include ``coalition formation'' \cite{kahan2014theories} and ``boundary activation'' \cite{tilly2004social} but these tend to focus on the building of group solidarity, not the initial person-to-person identification. \final{We also note that while the connotations of ``allyship'' often imply support across power asymmetries, our usage is more general and does not assume differences in power or influence.}}} to succeed. But, disclosure is risky because the exact conditions under which these situations arise heighten concerns of exposure and retaliation~\cite{ayres2012information,feldblum2016select}. That is, one cannot know whom to trust about an issue until one discloses being affected by it; yet one cannot disclose before knowing whom to trust.

{Some of the most helpful information in navigating problematic power dynamics can come from individuals who share similar traits regarding the context, e.g., women working with the same manager.
How can we facilitate the flow of information in such situations, so that individuals who need it most can discover allies while maintaining one's anonymity?} We consider this problem in the context of PhD students experiencing advising challenges. Because the advising relationship is fraught with power dynamics and ripe for exploitation~\cite{swazey1993ethical,lofstrom2020ethics,cohen2022abuse,martin2013countering}, ally discovery is critical for enabling students to connect with peers who might offer advice or partnership toward remedial action. Yet, research shows that ally discovery within the same PhD program can be difficult~\cite{friedensen2023power,golde2005role}.

% To build a system to facilitate discussions among PhD students and enable ally discovery, our study involved three phases. 
We conducted a three-phase study to build and evaluate a system that facilitates discussions among PhD students to enable ally discovery.
In Phase 1, we conducted 19 interviews across nine R1 institutions, and found that participants had a strong desire for a social platform where they can share information about advising dynamics. Many pointed out advising challenges is a sensitive topic that is not easy to talk about, even with peers in the department. Furthermore, participants emphasized wanting to have control over who can see their posts based on a rich set of contextual factors, which existing social media platforms do not provide \cite{kim2025trust}.

In Phase 2, we combined insights from Phase 1 interviews, prior work, and the affirmative consent framework~\cite{im2021yes}, to develop Moa (모아),\footnote{``Moa'' is a transliteration of ``모아,'' which is the conjugated form of the Korean verb ``모으다'' (to gather).} a fully functioning social media platform. Beyond features of existing social platforms for sensitive discussion, Moa includes a novel concept we call the \emph{consent boundary}. Existing social media platforms tend to offer brittle controls for audience configuration \cite{kim2025trust,zhang2020configuring}, or they provide mechanisms for selecting individual recipients on a one-on-one basis, which requires knowing users’ identities in advance. In contrast, consent boundaries allow users to demarcate who can view each post or comment based on dimensions such as social identity, lived experience, and interpersonal relationships, while maintaining all users' anonymity---the sender does not need to know the recipients' identities, nor must the sender reveal their own.
% They were designed in response to the principles of affirmative consent~\cite{im2021yes} and Phase 1 findings---participants emphasized wanting to have control over who can see their posts based on a rich set of contextual factors, which existing social media platforms do not provide \cite{kim2025trust}.

In Phase 3, we evaluated Moa in a 3-week field study with 47 PhD students from two departments at a public R1 university. Through system logging, a survey, and interviews, {we found that overall, Moa met its goal of enabling ally discovery among PhD students. Out of 47 users, 31 posted or commented on advising-related topics, some of which were sensitive, such as whether to switch advisors. In post-hoc interviews, many commented on their ongoing interest in Moa; a few expressed interest in directly contacting other PhD students on the system or even meeting someone in person.} 

\final{Out of 31 users who posted or commented, seven (22.6\%) used consent boundaries, suggesting that while consent boundaries were not a critical element for ally discovery, they were useful to some. Those users who used consent boundaries had various reasons for doing so—for some, it was to restrict the audience size and reduce privacy risks; others used them to proactively curate an audience (as opposed to defensively setting a boundary) in order to reach those who were sympathetic, could provide useful input, or might find the content helpful.}
% Participants cited various reasons for using consent boundaries—some aimed to prevent their own or their advisor’s identity from being revealed. However, not all were tied to reducing privacy risks; others sought to curate an audience that was more likely to be sympathetic or useful to the discussion. 
% {Most participants decided not to show their consent boundaries, as they were focused on audience configuration, rather than sharing parts of their identity to other users.}

While Moa is a basic social media prototype, its uniqueness lies in the combination of features designed to enable ally discovery. In all, this work {makes} the following contributions:
\begin{itemize}
    \item A recipe of socio-technical features for developing social platforms that facilitate ally discovery, informed by both our study and prior research. 
    \item Conceptualization and instantiation of consent boundaries as a way to address the issue of coarse audience configuration on social media systems.
    \item Reflections on \final{taking a consentful approach to developing systems, as compared to taking a privacy-centered approach}.
\end{itemize}

\section{Related Work}
We first review research on problematic power dynamics. Because ally discovery requires disclosure for others to take notice, we then discuss literature on audience configuration and identity management—key factors that impact users' self-disclosures. We next give an overview of research on consent. Lastly, we discuss challenges in PhD advising relationships.

\subsection{Studying Problematic Power Dynamics or Organizational Justice}
%(1) HCI (and other computing) has addressed collective action in abuse of power situations,
Rankism refers to ``abusive, discriminatory, and/or exploitative behavior towards people because of their rank in a particular hierarchy'' \cite{fullerrankism}. There is little work that studies how prevalent rankism is across contexts. For example, while research in both North American and European contexts show at least 10--15\% of the workforce has been exposed to hostile, aggressive, or abusive behaviors, it is still unclear what percentage of instances are perpetuated by people in power~\cite{branch2013workplace,zapf2020empirical}.
% Yet, news reports clearly show abuse of power is a problem that is hard to identify in a time-sensitive manner. For example, faculty’s academic misconduct, which often includes exerting pressure on graduate students to be part of it, takes a long time to be surfaced \cite{o2024embattled,kaiswer2023stanford}.

While not focused on problematic power dynamics in interpersonal relationships, a few HCI systems are relevant for enabling sensitive discussions in organizational contexts. For example, Abdulgalimov et al. \cite{abdulgalimov2020designing} built OurVoice, a platform that helps employees discuss issues in the workplace. A deployment study showed that anonymity, strict identity verification, and moderation were important in enabling sensitive discussions about organizational justice (e.g., gender pay) among employees \cite{abdulgalimov2020designing}. There has also been research on systems for collective action, such as Dynamo, a platform for Amazon Mechanical Turk workers to organize \cite{salehi2015we}. 
% The authors uncovered stages of collective organizing, such as stalling and friction, by developing the platform and engaging with Turkers \cite{salehi2015we}.

Compared to such prior work, we focus on a more interpersonal context of individuals navigating problematic power dynamics within an organization. Research shows that it is extremely hard to be the first person to disclose some kind of information for potential allies to notice in such situations~\cite{ayres2012information}. This is fundamentally related to the difficulty of building trust---trust is about the willingness to be vulnerable with someone in a situation that assumes some kind of risk~\cite{mayer1995integrative,schoorman2007integrative}.
% Reasons include fear of experiencing further trauma \cite{liu2018whispers}, damage to reputation \cite{ayres2012information}, and extreme uncertainty as to whether others in the same situation would also come forward~\cite{cheng2022reporting}. 

\subsection{Audience Configuration and Identity Management Online}
Audience configuration and identity management are key factors that impact self-disclosures on social media systems~\cite{ma2016anonymity}. 
% Many social media platforms rely on anonymity, a primary mechanism for concealing one's identity, to support sensitive discussions~\cite{leavitt2015throwaway,kim2017blind}. While research shows that audience configuration also plays a role in sensitive disclosures~\cite{ma2016anonymity}, less work has focused on enabling users to configure their audience more flexibly. Researchers have pointed out that this is likely due to the status quo of how social media platforms are designed, which often provide coarse-grained controls~\cite{zhang2020configuring,paci2018survey}.

\subsubsection{Audience Configuration}
While research shows that audience configuration plays a role in sensitive disclosures~\cite{ma2016anonymity}, less work has focused on enabling users to customize their audience more flexibly. Researchers have pointed out that this is likely due to the status quo of how social media platforms are designed, which often provide coarse-grained controls~\cite{zhang2020configuring,paci2018survey}.

\paragraph{Social media structure.} Fundamentally, the form of a social media system impacts how users configure their audience. Zhang et al.'s Form-From framework \cite{zhang2024form} classifies three ways social media systems distribute content: \textit{networks}---users create connections (e.g., friending, following) to exchange information,
\textit{spaces}---all users within a shared environment (e.g., channels, email threads) see the same information, and
\textit{commons}---users can receive content from anyone, without having to follow a user or join a space.

Many studies on audience management on social media have focused primarily on \textit{networks}~\cite{wang2016modeling,collins1994self,bernstein2013quantifying}. The coexistence of different types of social ties (e.g., family, co-workers) on many platforms, which is often called the ``context collapse'', makes it difficult for users to carve out a precise audience \cite{davis2014context,marwick2011tweet,vitak2012impact}. A study showed that Facebook users actually tend to underestimate their posts' audience size~\cite{bernstein2013quantifying}. 
% For example, Ma et al. \cite{ma2016anonymity} found that an audience based on social ties increased people's self-disclosure than one based on physical proximity.\footnote{For example, Yik Yak lets users post to an audience of people within a 5-mile radius.} 

Some platforms that fall under \textit{commons} recommend content and accounts by algorithmically inferring users' interests, such as Snapchat's Discover~\cite{jones2023train,kim2025trust}. However, research shows that users perceive them to be still ineffective in identifying the right audience~\cite{kim2025trust}.

There is a rich line of work on audience configuration in emails, which has the form of a \textit{space}. Similar to social media platforms, people actively consider social affiliations when configuring an audience using the To and Cc fields~\cite{zhang2020configuring}. Other studies have examined the impact of Ccing others in workplaces~\cite{haesevoets2021transparency,machili2019snowball,skovholt2006email}, as well as how people decide whether to reply or reply all~\cite{grandhi2016reply}. However, emails require users to know who their audience is and how to reach them.

In this work, rather than being constrained by a predefined network, space, or algorithmic configuration, we ask: ``How does each individual user want to form an audience \textit{per post or comment}, when they do not know the audience's exact identity?'' 

\paragraph{Access and privacy controls}
Even with the same underlying structure, platforms could differ in respect to what are often called access, privacy, or permission controls \cite{powers2002privacy,paci2018survey}. Privacy researchers tend to classify these controls into two categories: interpersonal and institutional \cite{lowens2025misalignments}. Interpersonal controls regulate how other users access one’s personal information. Settings for audience curation fall under this category, and many general platforms offer coarse options, such as ``Public'', ``Friends'', and ``Close Friends'' \cite{lowens2025misalignments,wisniewski2012fighting}. Institutional privacy controls manage what platforms collect about users, particularly in the context of online behavioral advertising \cite{im2023less,lowens2025misalignments}. 
% On more specific systems, access controls serve distinct purposes, such as for private collaboration on crowdsourcing systems \cite{yuan2021sleuthtalk}.

While the usable privacy community has advanced the design of access controls for users (e.g., \cite{im2023less,schaub2020usable,yao2019defending,habib2021toggles,feng2021design}), much of the work tends to focus on increasing the controls' usability, such as ensuring that they are easy to find \cite{im2023less}. Notably, Feng et al.~\cite{feng2021design} proposed a design space for privacy controls, introducing five dimensions—type, functionality, timing, channel, and modality.

The social computing community has particularly argued for more radical designs for audience configuration~\cite{kim2025trust,paci2018survey,im2021yes,baughan2024supporting,kim2024respect,page2019pragmatic}. Most recently, Kim et al. \cite{kim2025trust} articulated the value of viewing privacy management as a trust-driven process, and suggested concrete designs based on such reframing, such as contextual audience segmentation, intentional engagement signaling, and guided disclosure.

\subsubsection{Identity Management}
Anonymity is known to help with sensitive disclosures because the lack of identifiable ties lowers risks \cite{ma2016anonymity}. Because of this, major social platforms rely on anonymity for posting sensitive information. For example, users actively use throwaway accounts on Reddit \cite{leavitt2015throwaway,ammari2019self}. Blind, a platform launched in South Korea, enables anonymous sharing about employers, verifying users via corporate email addresses~\cite{kim2017blind,whittaker2018blind}.\footnote{\url{https://www.teamblind.com/}}
% Whisper is another anonymous platform, which markets itself as ``the best place to discover secrets around you.''\footnote{\url{https://whisper.sh/}}

HCI system researchers have explored how anonymity, coupled with moderation, helps with sensitive conversations, along with new identity management mechanisms beyond anonymity. For instance, OurVoice supports sensitive workplace discussions by combining anonymity, strict identity verification, and moderation~\cite{abdulgalimov2020designing}. {Meronymity, a set of traits that provide information about a person's identity without revealing it completely, helped junior scholars participate in academic discourse on Twitter, mitigating the impact of social hierarchies~\cite{soliman2024mitigating}. The goal of meronymity is to enable users to present themselves more flexibly to give sufficient context to others, while preserving an adequate level of privacy~\cite{soliman2024mitigating}.}

However, identity management does not fully resolve users' uncertainty around disclosing information or connecting with people online. For example, research shows that anonymity does not always translate into meaningful social connections, because it does not ensure empathetic responses—an essential factor in sharing sensitive information~\cite{ma2016anonymity}. This is likely one reason why many discussions using throwaway accounts end as one-off disclosures~\cite{leavitt2015throwaway}. {And, even for participants who used meronymity, some expressed hesitation or fear of revealing too much about themselves~\cite{soliman2024mitigating}, which shows that audience control is a key factor for enabling sensitive discussions.}\newline

\noindent {Overall, no work has yet instantiated fluid and contextual audience configuration on a social platform. While our work started off as developing a system to support sensitive discussions around power dynamics, we address this gap in audience configuration as part of our contributions. }

% \noindent {In this work, we propose a new concept called consent boundaries, that combine audience control and identity management in a novel way. The core idea of consent boundaries is to let users precisely decide who they want a post or comment to be shown to, while they do not know the identities of other users. And, at the same time, users can choose whether to show the perimeters of their consent boundary to the selected audience. The ability to partially disclose information about oneself without revealing their identity resembles meronymity~\cite{soliman2024mitigating}; yet, our focus is more on audience control, with meronymity being a natural coupling to the feature, to enable more information-sharing between the poster and the audience (if the poster wants to).} 

\subsection{Consent}
In essence, consent is a grant of rights to another person to initiate some kind of interaction. This is especially important for high-risk contexts, such as sexual interactions \cite{muehlenhard2016complexities} and medical procedures \cite{kinnersley2013interventions}. For example, in sexual contexts, asking for consent acknowledges that one person can only engage in sex if the other party explicitly grants that right. This is why consent clearly explains why sexual violence is morally wrong—it emphasizes respecting each person's autonomy and decision-making power~\cite{friedman2019yes}.

While consent is a powerful concept, defining it has not been straightforward. There have been debates on whether to view consent as a subjective or objective phenomenon \cite{halley2016move,bryden1999redefining,wertheimer1999consent}. \textit{Subjective consent} is based on one's internal sense of having granted permission, while \textit{performative consent} is an indication of agreement based on behavioral signs or explicit verbalization \cite{halley2016move,hurd1996moral,beres2007spontaneous}.
% In the context of sexual interactions, many scholars have argued for performative consent \cite{beres2007spontaneous}. But, performative consent can differ from subjective consent, especially if performative consent is given under pressure or deception \cite{beres2004sexual,beres2007spontaneous,mcgregor1996she}. Yet, without some overt signals, parties might never know when consent is truly granted \cite{beres2007spontaneous,muehlenhard1996complexities}. Furthermore, solely emphasizing a person's subjective state can lead to unnecessary scrutiny of the mental state of survivors of sexual violence \cite{kazan1998sexual}. 

\textit{Affirmative consent} is a form of consent that values both subjective consent and performative consent. In essence, affirmative consent is an explicit indication of voluntariness and enthusiasm before an interaction is initiated \cite{friedman2019yes}. Building on Una Lee's work on ``consentful technologies'' \cite{lee2017consent}, Im et al. \cite{im2021yes} argued that affirmative consent is \textit{voluntary}, \textit{informed}, \textit{specific}, \textit{revertible}, and \textit{unburdensome}, and to develop social computing systems based on these five properties.

% \subsubsection{Consent in Interactions with Systems}
In the field of computing, consent is often viewed as a concept for data control within the \textit{notice and choice} model, in which users are given notice about how their information might be collected and used, and offered choices about it  \cite{obar2020biggest}. 
However, many scholars have criticized the model because it has failed to protect people’s privacy in practice \cite{cate2006failure,madden2017privacy,nissenbaum2011contextual,reidenberg2015disagreeable,solove2012introduction}. In this work, instead of being confined to the notice and choice model, we view consent as a fundamental concept that give users the agency to carve out their online interactions. 
% In particular, the affirmative consent principles of specificity, revertibility, and unburdensomeness \cite{im2021yes} enable more granular permissions.

\subsection{Challenges in PhD Advising Relationships} 

PhD advisors play a significant role in their students' mental health \cite{peluso2011depression,levecque2017work,friedensen2023power,breen2024breaking} or success in the program \cite{golde2005role}. In a worldwide survey by Nature, 23\% of PhD student participants answered that they would switch advisors, if given the chance \cite{woolston2017graduate}. Another study from a European institution shows that 24\% of PhD students perceived they have experienced some kind of abuse from faculty \cite{jacob2018aveth}. In HCI, a qualitative study showed that graduate students cite advising relationships as one main cause of stress in graduate school \cite{robledo2023we}. Other research shows the range of advisor behavior that can cause harm: incompetent advising, sudden dismissal of advisee, forcing advisor's views, exploitation, bullying, encouraging fraud, authorship issues, misappropriating a student's work, harassment, abuse, and dual relationships (e.g., friendship, romantic relationship) \cite{lofstrom2020ethics,cohen2022abuse,martin2013countering,barnes2009nature}. 
Research also shows that even when PhD students realize they are experiencing harm from their advisor, they have few guidelines for what to do next \cite{becerra2021does,golde2005role}.

\section{Phase 1: Formative Study}
We conducted two rounds of semi-structured interviews with 19 PhD students in the United States.  Our study was exempt-approved by our institution's Institutional Review Board.
% We conducted two rounds of interviews that resulted in 19 interviews across 10 PhD programs at 9 institutions in the United States.

\subsection{Methodology}

% {{Dramatically cut the methodology section here. There is a lot of unnecessary detail, even for a regular Methodology section, and here, you really just want to quickly get to the findings. The main things to mention are 2 rounds of interviews; total number of interviewees from where; and semi-structured interviews + think-aloud protocol with low-fidelity designs; focus of the interviews. I don't think you need more than a couple of paragraphs.}

\subsubsection{Participant Recruitment and Study Protocol}
In the first round, we recruited 10 participants from 2 PhD programs within an R1 institution by contacting the student mailing lists.
For the second round, we recruited 9 PhD students from 8 R1 institutions by publicizing the study on X and LinkedIn. 
We noticed that a few participants self-disclosed their experiences to an open-ended question in the sign-up form, and prioritized recruiting them, while accounting for year in PhD and PhD program. Participants were compensated \$20/hour.
All participants were in computing-related PhD programs, ranging from small (fewer than 40 students) to large (over 200 students).
The average PhD year for first-round participants was 3.8, and 2.9 for the second round.

We showed participants a hypothetical scenario of a PhD student whose advisor persistently requested them to work on non-research tasks (see Supplementary Materials). We asked participants what they would do in the situation. Then, we presented participants with a high-level description of a system that lets students converse regarding advising and asked for their thoughts. Next, we showed participants low-fidelity designs of such a system, and asked them to think aloud. The interviews lasted between 1 to 1.5 hours.
% We also did not ask about the participant's experiences with their current or prior advisors per our IRB-approved protocol.
% but some participants brought up the harms and challenges they experienced in relationships with advisors or other faculty. 
% However, we do not report the exact number of how many participants brought up such cases per our IRB-approved protocol.

\subsubsection{Initial Designs} \label{initial-designs}
First, we considered a posting and commenting format (Figure \ref{formative-study-designs}). However, the design was not a public forum---we wanted to give users control over each post's audience, and considered two criteria: 1) types of PhD advising-related challenges and 2) faculty names. 
% A major goal of the interviews was understanding whether such criteria were useful, and if there were others that PhD students wanted. 
We also included a messaging feature because we hypothesized it would be helpful for PhD students to privately ask and answer sensitive questions.

% \begin{figure*}[t]
%   \centering
% \includegraphics[width=0.6\linewidth]{figures/moa-1.png}
%   \caption{Students can anonymously write their experiences on Moa. At the same time, the interface should help them write with some level of specificity.}
%   \label{fig:overview-moa}
% \end{figure*}

% \begin{figure*}[t]
%   \centering
% \includegraphics[width=0.6\linewidth]{figures/moa-2.png}
%   \caption{We envision that using the system, students can indicate \textit{who} problematic faculty are, with varying levels of anonymity.}
%   \label{fig:overview-moa}
% \end{figure*}

% \begin{figure*}[t]
%   \centering
%   \includegraphics[width=0.6\linewidth]{figures/moa-5.png}
%   \caption{Students can connect with each other using Moa's messaging feature. The key here, is to enable connections without students knowing each other's identity.}
%   \label{fig:moa-5}
% \end{figure*}

% \begin{figure*}[h]
%   \centering
% \includegraphics[width=0.6\linewidth]{figures/moa-4.png}
%   \caption{Students can see other students' posts. This mockup shows a student describing a serious incident of their advisor asking them to falsify research data.}
%   \label{fig:overview-moa}
% \end{figure*}

\subsubsection{Data Management and Analysis}
The first author manually corrected the auto-generated transcripts. For sensitive personal experiences, we either redacted it or asked participants about inclusion. 
% Two participants requested further redactions, to which the first author revised the transcript accordingly. 
The first author conducted inductive coding and periodically discussed the themes with the second author.

\subsection{Phase 1 Findings}

We report the most salient findings from the Phase 1 interviews.

\subsubsection{Needs and challenges in connecting with other PhD students about advising}
Almost all of the participants expressed interest in using a system that connects PhD students regarding advising relationships. Even the one participant (P12) who had mixed feelings acknowledged the sytem's value, though they felt it did not address the root issue of faculty oversight: \textit{``I feel like overall, there just needs to be more oversight [of] faculty and PhD programs. [...] And I think this [system] could be a really great place for students to start doing that bottom-up type of accountability thing.''}

When asked what they desired the most, almost all participants answered they wanted \textit{validation} of their experiences. They wanted to know whether their interactions with their advisor were ``normal'' or problematic. For example, P17 said \textit{``Just knowing that other people's experiences are either the same or different gives you courage to know that it's OK or it's not OK. It helps you figure out what's a normal experience.''} 
% Similarly, P15 commented \textit{``I think is already nice to know, like people experienced it or are currently experiencing it. I think that's a lot to make it feel like it's not in your head, you know, like it makes you not feel crazy.''}

Interviews also revealed why PhD students struggle to seek validation and advice about their advising experiences. Many participants found it was difficult to discuss their experiences with peers or mentors in their established social circles (e.g., research group) due to concerns about power imbalances---a key reason they wanted a system that could facilitate conversations beyond those circles. This contrasts with prior literature emphasizing the role of social ties in sensitive disclosures \cite{ma2016anonymity}. For example, some participants avoided discussing their advisor with labmates out of fear of retaliation. P16 explained \textit{``I think it can be hard to talk to other PhD students in your lab or your department because you don't want that information to get out and get back to your advisor.''}

Similarly, some participants who had switched advisors avoided discussing their experiences within their close networks due to concerns about repercussions and damage to their reputation. For example, a few participants said they avoided disclosing advising experiences to students in the same program to avoid burning bridges with former advisors. P15 said \textit{``If you've been through a negative advisor experience, I feel like not everyone is willing to be transparent about that with, say, people who reach out to them asking about how their experience in a particular group was so. And it's because we know there are power hierarchies in academia.''}

Additionally, some participants noted that even within the same institution, connecting with others who had faced the same kind of challenges could be difficult. P6 said \textit{``...let’s say
it’s the same department, but [the PhD student is in] some other lab,
and I’ve never interacted with the person before. The chances of us
connecting over that experience is very low.''} Similarly, P13 observed \textit{``So if they [specific PhD students] are not in the exact same scenario, their suggestions would be different.}'

% \begin{quote}
%     Jane: And so I want to focus on this particular context in this situation, would you use this system at all? Like, would you write something at all?

%     Participant: Yeah, I would! I would say like hi, like, my professor keeps doing XYZ thing. Isn't it normal? [Chuckles]  
% \end{quote}

% P19: I just had an idea flutter through my mind. I don't actually know if I agree with it or not, but like group chats, but maybe that's just like people commenting on each other's posts instead of having one-on-one. Kind of like forum style one can express their what they have to say about the situation, if they meet the criteria.

% This was less of a pattern in the first batch, probably because the participants all knew the first author as they are from the same institution. Two participants (2/10) mentioned they would be open to ``grabbing coffee,'' but even for these two, they all emphasized that trust was a requirement. 

\subsubsection{Form of the platform: Consent and control in audience setting} \label{not-chat}
Regarding the platform structure, many participants wanted a form that enabled lightweight, distant interactions before mutual trust developed. Broadly, this meant they favored a post-and-comment structure over a messaging system. We initially hypothesized that PhD students would seek deeper connections when supporting one another, such as through a direct messaging. However, most participants did not want dyadic, deep interactions due to the uncertainty of the other person’s identity and the context's sensitivity. Others noted that 1:1 messaging created an obligation to reply. As P16 explained: \textit{``I think it would be easy [to write a post], especially in a forum setting where maybe I wasn't like obliged to respond back to people because other people are chiming in as well.''}

At the same time, participants wanted something different from a traditional forum—many preferred the ability to set specific criteria for who could access their posts. Participants reacted positively to proposed audience control features and recognized that this structure differed from existing forums (see Section \ref{initial-designs}). For instance, P19 said \textit{``Kind of like forum style one can express their what they have to say about the situation, if they meet the criteria.''} One participant (P16) explicitly used the word ``consent'' when describing how the discussions could have some kind of access control: \textit{``I guess people would have to like consent to this, but if you could be directed to a group with people that have been having similar issues, and you could see what people have been saying in threads of conversations, and then you could join in.''}

Participants' reasons for wanting granular levels of control included privacy concerns and fear of repercussions. For instance, P11 said: \textit{``I would have to feel more comfortable on the platform without revealing identity stuff. And I would want to know that they're not like in the same lab as me or share the same advisor, you know.''} 
Another reason was the importance of gradually developing trust before disclosing more information—something existing forums fail to support because they tend to be public.
For example, P12 said \textit{``I think that's one of the reasons why like platforms like Reddit don't do so well in terms of forming like strong social support networks because everything is too public, if that makes sense, like you need some sense of like trust and privacy and intimacy to have to start building those relationship blocks to actually support someone.''}

\subsubsection{Criteria for deciding whom to interact with} \label{phase1-boundary-criteria}
% All participants reacted positively to the feature for selecting specific advising challenges to carve out an audience (Section \ref{initial-designs}).
Participants found our proposed criteria of advising-related challenges and faculty names to be helpful (see Section \ref{initial-designs}). The interviews also revealed additional criteria that participants suggested.

One important finding was that some participants naturally brought up identities—such as gender, race, and being an international student—without being prompted. (We did not initially include social identities in our list of criteria.) For instance, P16 said \textit{``If I had one [difficulty in advising] with sexism, then maybe I'd want to limit it to like other female identifying students or maybe not men. So yeah, maybe depending on the issue, there would be other like demographic information like that.''} 
Similarly, P18 highlighted the challenges international students face: \textit{``...as international students, the education system here is so different from the system that we grew up in. We have no idea for resources for harassment, you know, threats or anything.''}

Many participants also wanted the ability to filter by PhD programs or even specific labs (e.g., avoiding labmates). For instance, P19 said \textit{``I really would not want to have that conversation with them [labmates] over this platform [...] But maybe that's part of the criteria, right, like you say `Oh I don't want anyone else who is being advised by X.' ''}

% A few participants noted that they would only like to connect with PhD students who have experienced the same challenge \textit{and} successfully resolved it. For instance, P15 said \textit{``I would need to talk to someone who's been through this and has kind of resolved the situation. I may not want to speak to someone who is going through it right now.''}

Some participants said they are fine with posting publicly. Their main reason was to get advice from a wide audience.

\subsubsection{Preference towards anonymous interactions} \label{shallow-interactions}
Nearly all participants preferred anonymous connections. When asked whether they would ever reveal their identity to another user on the platform, most participants said they would not. Many thought anonymity was important to reduce the risks of sharing sensitive information. P19 said \textit{``You can be more open with people if you're anonymous.''} Others noted that anonymity helps mitigate bias. For instance, P13 said \textit{``...because if I reveal my identity, other people think I'm some kind of vulnerable thing or I cannot progress on my research so that I want to quit.''}

At the same time, participants acknowledged that the desire for anonymity gradually weakens as trust develops. For instance, P19 said \textit{``I don't think identity has to be all or nothing, it can be parts of it and I would do it as the other person was opening up, like you would actually do, almost in a conversation.''}

% \subsubsection{Security and credibility}

\subsubsection{Trust between users and the system creators} \label{user-to-system-consent}
Many participants asked questions about the system creators' affiliation and their access to users' data. For instance, P15 said \textit{``I think the biggest thing I want is some kind of transparent security and privacy labels or like who is going to have access to this database? And I want to know that this is very much like student organized and not something coming from faculty.''}
Some participants commented on the importance of knowing the system creators' motivation. P7 said \textit{``Maybe if there was a picture of you [first author] on the home page and it says yeah, man, like I had some fucked up shit happen to me, that might actually be persuasive to me. There are people who made this, not just because they wanted to write a paper, but because of what happened to them in their PhD.''}
% Relatedly, when asked about whether they would feel comfortable having a human moderator being involved in connecting PhD students, many participants preferred algorithms or some kind of combination of algorithms and human moderation.

\subsubsection{Preventing abuse}
Many participants desired strong abuse prevention measures, so that they would know for sure a user is a student, and not a faculty member. 
Participants were also aware that students could abuse the system, such as by spreading false information about faculty. For example, P10 said \textit{``Sometimes the abuse can come from the advisee’s side, although that is rare. So, they [PhD students] need to be fair.''} 
When informed that we plan on moderating the system, many participants emphasized wanting transparency behind moderation \cite{jhaver2019does}.

\subsubsection{Summary}
Based on the interview findings, we organized our design requirements as the following. The platform should:
\begin{itemize}
    \item REQ1: Follow a post-and-comment structure.
    \item REQ2: Provide consentful controls for audience setting with respect to various contextual information, such as social identities, advising experiences, affiliation, and advising status.
    \item REQ3: Ensure interactions remain anonymous; minimize risks of revealing one's identity.
    \item REQ4: Have strong privacy and security measures.
    \item REQ5: Include moderation and community guidelines to prevent potential abuse.
\end{itemize}

\begin{figure*}[t!]
\centering
\includegraphics[width=0.9\textwidth]{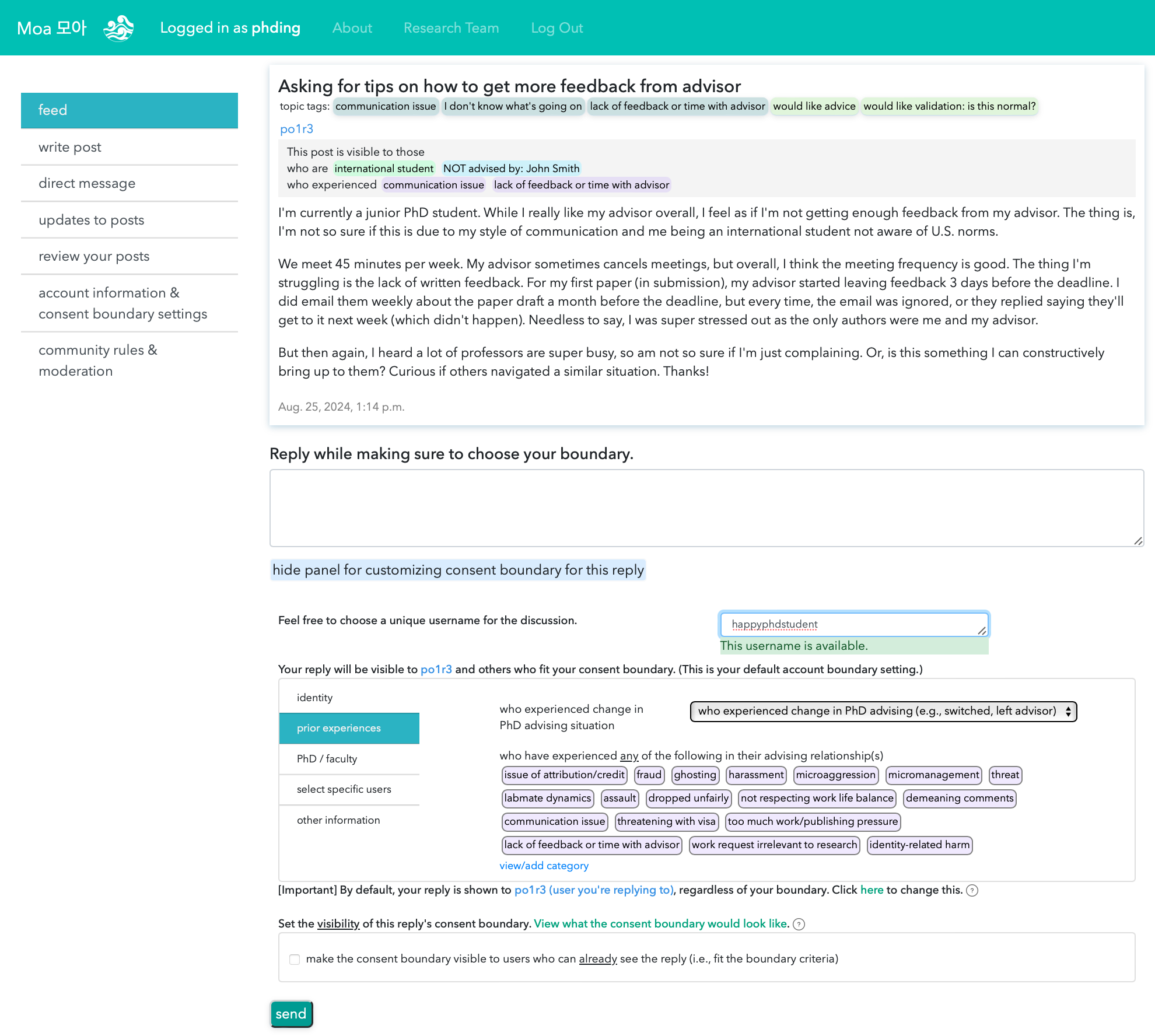}
\caption{A user (@po1r3) set their post to be visible to those who are international PhD students, not advised by John Smith, and have experienced communication issues or lack of feedback. A user (@phding) is about to reply while creating a different username (``happyphdstudent''). @phding can see the post's consent boundary because  @po1r3 opted in to showing it to those who can already see the post.}
\label{fig:overview2}
\end{figure*}

\section{Phase 2: Design of Moa}

Moa is designed according to insights gained from Phase 1, previously known features that enable sensitive discussion, and the principles of affirmative consent \cite{im2021yes}---the value of the last being reinforced by the Phase 1 interviews. 
% Below, we begin with a description of Moa's overall design, and then in Section~\ref{consent-boundary}, discuss consent boundaries.

\subsection{Overview of Moa}
Like many social platforms, on Moa, users can write text-based posts and comment in response to posts (REQ1). 
% Each feature is not necessarily novel, but their combination intended to encourage sensitive conversation makes Moa unique.
% Moa is a basic social media system that allows users to communicate with one another online. 
% Like many social media platforms, users can write text-based posts and comment in response to posts. 
% What makes Moa unique is the combination of features intended to encourage sensitive conversation. 

\subsubsection{Not Network- or Space-Based; No Profile Page} \label{no-network-space}
Social media platforms are typically structured around a predefined network (e.g., followers) or space (e.g., group) \cite{zhang2024form}. In contrast, Moa focuses on the question: ``How does each user want to design their audience per post or comment?'' As a result, Moa does not implement any network model nor does it include user profile pages (REQ3).\footnote{Among existing platforms, Whisper also does not provide profile pages.}
Furthermore, instead of letting users have access to the same content if they are in the same space, Moa's users see different sets of posts (and comments) depending on their own and others’ consent boundaries (Section \ref{consent-boundary}). In terms of the Form-From matrix \cite{zhang2024form}, we believe Moa is closest to a Threaded Space, but consent boundaries considerably weaken both the threading and the space.

%as is explained during the sign-up process and on Moa's ``about'' page. The first author checked whether the user was a PhD student enrolled in one of the two PhD programs by using the PhD program student lists. They also skimmed their self-declared identity traits, such as gender and race, to ensure they were accurate. A user can only access Moa's content after they verified their email account and the moderator has verified their identity.

\subsubsection{Strict Identity Verification} \label{identity-verify}
Because all Phase 1 participants strongly emphasized that they wanted a space of only PhD students, Moa has a strict identity verification process (REQ4). Users can only sign up using an institutional email address; email domains are verified by the system (Supplementary Materials).\footnote{For users who did not feel comfortable leaving their institution email address in the database, Moa lets them change to a personal one after signing up.} Furthermore, the moderator (the first author) reviews every sign-up, verifies email validity, enrollment in one of the two PhD programs, and personal traits (to the extent that can be publicly verified) as entered by the prospective user. 

\subsubsection{Multiple Usernames} 
Phase 1 participants emphasized preserving users' anonymity on Moa. A few participants said they wished there were multiple user identifiers on the system, so that it would be hard for others to infer their identity. 
Based on this finding and affirmative consent's \textit{specific} principle \cite{im2021yes}, we let users create different usernames per post or discussion (REQ3; Figures \ref{fig:overview2}). Once a user takes a username, it cannot be used by others, even if they are in a different thread.
Compared to other platforms, Moa provides this feature directly in the posting and commenting interface (\textit{unburdensome}), which is not the default for major social media platforms.\footnote{On Blind, a user has to go to their account setting to change their username. Since August 2023, Blind allows users to change their username once per day. Previously, it was allowed 5 times per month. \url{https://www.teamblind.com/post/New-More-username-changes-7CXLWNB4}}

{To prevent users from using multiple usernames to add weight to their point, once a user starts or participates in a thread, they can only use the name that they initially used. The username field becomes deactivated, so that the user can no longer make modifications to it.}

\subsubsection{Community Rules and Moderation}
Based on the Phase 1 findings and prior research \cite{abdulgalimov2020designing,seering2017shaping}, Moa emphasizes community rules on a linked page from the navigation bar (REQ5), which are: 1) Be respectful and kind to others, 2) What is said on Moa, stays on Moa, 3) Do not spread false information, 4) Do not use discriminatory language, 5) Do not harass or attack others. Moa is moderated by the first author because adding others would heighten users' privacy concerns. 
However, theoretically, Moa could be extended to recruit other moderators.

\subsubsection{Notifications}
Whenever a post is created, users within the post's consent boundary are notified via email.
Moa also notifies all users in a thread whenever a user adds a new comment. For both, the email shows the first 20 words of the post/comment. 
% The notifications also appear in each user's updates page.

\subsection{Consent Boundaries}%\subsection{Marking a Consent Boundary Per Piece of Content Based on Experiences and Identities} 
\label{consent-boundary}

Moa lets users mark their \textit{consent boundary}, which means users can set an audience per post or comment based on a rich set of dimensions (REQ2; \textit{voluntary}, \textit{specific}). 
% For example, a user could write a post that is only visible to other users who have experienced the same type of advising challenge (Figure \ref{fig:overview2}).

\subsubsection{Flow of setting a consent boundary}
A user can set a consent boundary while writing a post (or comment) and save it to their account setting (Figure \ref{fig:overview2}). Or, a user can set a default, account-level consent boundary in the settings page. Whenever a user sets a  consent boundary for a post/comment, it does \textit{not} affect prior ones (\textit{specific}). When a user visits a post they wrote (or a thread they participated in), Moa pulls up the  recently used boundary for that post (\textit{unburdensome}).

\subsubsection{Dimensions of consent boundary} \label{consent-boundary-dimensions}
Below, we list possible consent boundary dimensions.

\begin{figure*}[t!]
    \centering
    \includegraphics[width=1\linewidth]{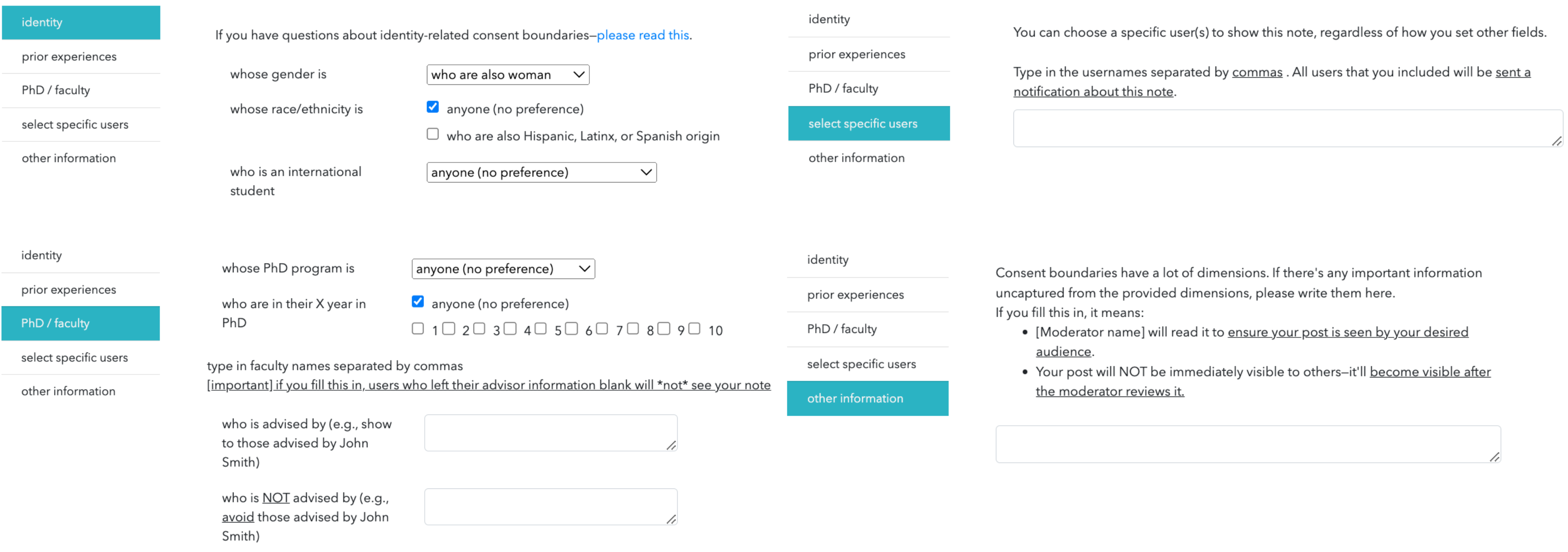}
    \caption{Different dimensions of consent boundary. The high-level categories included: identity, challenges experienced in advising relationships (Figure \ref{fig:overview2}), PhD program/faculty-related factors, specific users, and other information. We explain each category in Section \ref{consent-boundary-dimensions}.  }
\label{boundary-dimensions}
\end{figure*}

\paragraph{Identity}
Moa lets users show their post or comment to others who share their gender, race, or international student status.
We enabled users to select each category \textit{only} when they have the corresponding identity.\footnote{{We note that was a deliberate design choice among several options. For example, a platform could let users set boundaries that they are not themselves within, perhaps to seek opinions from groups they are not members of. However, that can open the door to misuse without additional safeguards, so for this work, we opted for the more conservative option.}} For example, only women can set their consent boundary so that their posts and comments are visible to women. We describe this more in Section \ref{declare-identity}.

\paragraph{Prior advising experiences}
Moa lets users show their post/comment to others who have experienced specific kinds of advising challenges (e.g., micromanagement, communication issue; Figure \ref{fig:overview2}). The categories of challenges are based on research on PhD advising \cite{cohen2022abuse,lofstrom2020ethics} and Phase 1 interviews (full list included in Supplementary Materials). We also linked a survey in the interface so that users could suggest new categories they desired (\textit{voluntary}). 
Users can also only show their post/comment to others who have experienced a change in their advising situation (Figure \ref{fig:overview2}).

\paragraph{PhD program and faculty names}
A user can make their post or comment only visible to PhD students from a specific PhD program (Figure \ref{boundary-dimensions}).
Moa also lets users avoid, or connect to, other users who are advised by certain faculty (Figure \ref{boundary-dimensions}).

\paragraph{Specific usernames} Moa lets users choose accounts they only want to show a post or comment to (Figure \ref{boundary-dimensions}). For comments, users are only able to select users who participated in the thread.
% Moa notifies the selected users via email.
% 

\paragraph{Other information}
Users can type in information uncaptured by provided dimensions in the ``other information'' textbox (Figure \ref{boundary-dimensions}). When a user uses this, the post or comment is delayed until the moderator reviews it and chooses the right audience.

\subsubsection{Declaration of identities and experiences} \label{declare-identity} 

To make consent boundaries work, Moa has to know users' identity traits and experiences. Users can voluntarily declare their traits for \textit{only} the system to know (\textit{voluntary}).
Users are encouraged to submit information when signing up, but they can go to their account page to change it (\textit{revertible}). 
To ensure a user does not falsify their account information after sign-up, the first author periodically checked for users' account setting changes.

\subsubsection{Consensually visualizing consent boundaries} \label{visualize-boundaries}
\begin{figure*}[t!]
    \centering
        \includegraphics[width=.8\linewidth]{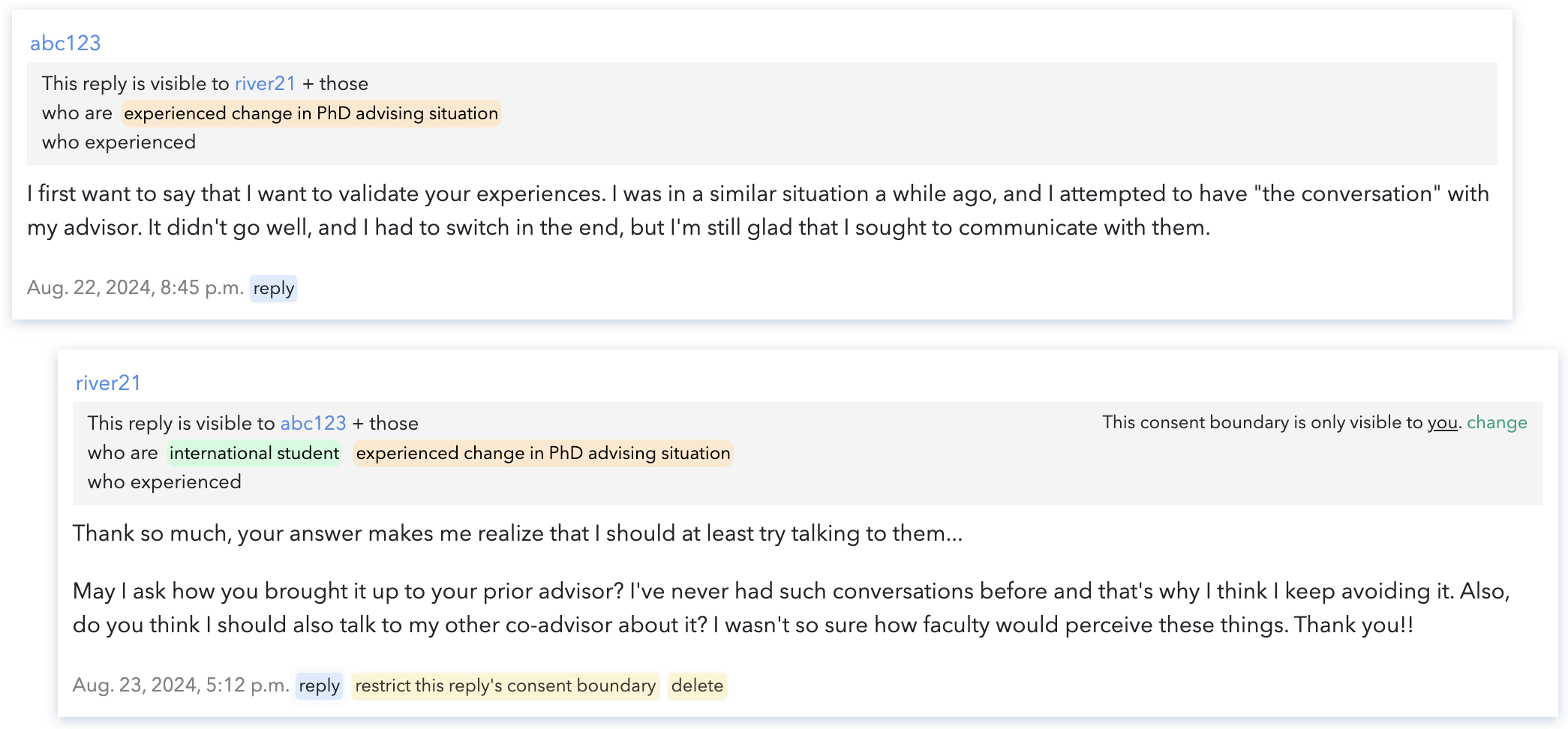}
 \caption{Users can also set a consent boundary for each comment. @abc123 shows their comment to users who have experienced a change in PhD advising, while @river21 limits their reply to @abc123 and international PhD students who experienced a change in advising situation. At the top left of @river21's comment, it says that the consent boundary is only visible to @river21.}
 \label{fig:overview3}
\end{figure*}

Moa lets users decide whether to show their consent boundary to others \textit{who can already see the post or comment} (\textit{voluntary}). {That is, by default, when users see a post with a boundary applied, it looks the same as public posts.} For example, in Figure \ref{fig:overview2}, the poster consented to showing their boundary. Those who have access to the post can understand that the poster is an international student, has experienced communication issues and lack of feedback from their advisor, and wants to avoid students who are advised by John Smith.
In contrast, if a user decides to not show their boundary, other users who have access can only see the content and username. Users can also change their boundary's visibility after posting or commenting (\textit{revertible}; Figure \ref{fig:overview3}).

\subsubsection{{Monotonically} decreasing branches} \label{mono-decrease-rule}

Moa applies the original post's boundary to all of its comments. This is because  research has shown that replies to a post can reveal its content, even if the post is deleted \cite{mondal2016forgetting}. 
A comment to another comment also inherits their boundary rules. 
% Of course, this is not communicated without the original poster or commenter's consent (\textit{voluntary}; Section \ref{visualize-boundaries}).

\subsubsection{Restricting a consent boundary}
Based on the affirmative consent's \textit{revertible} principle, Moa lets users \textit{restrict} consent boundaries (Supplementary Materials). Users who already replied to the post but no longer fit the new boundary will lose access to it (without any notification). 
% Users can also delete their posts and comments anytime.

\subsubsection{Not applying consent boundary rules to the user whom one is replying to}
There was one contradiction between the affirmative consent framework and how Moa's consent boundary was implemented. The framework is conservative and implies that consent boundary rules should be enforced without exception \cite{im2021yes}, including when a user replies to a post or comment. For example, let's say \textit{A} is a woman. \textit{A} set their default consent boundary so that their comments are visible to women. \textit{A} is about to reply to \textit{B}'s post. Based on the affirmative consent framework, Moa should enforce \textit{A}'s consent boundary to \textit{B} as well. This means \textit{B} cannot see \textit{A}'s comment if they did not indicate they are a woman in the account setting.

However, our pilot studies showed that this could be confusing. 
Thus, we changed it so that when a user leaves a reply, by default, Moa assumes they want to show their reply to the original poster. Users can always change the setting (see the bottom of Figure \ref{fig:overview2}).

\subsection{System Implementation and Pilot Studies} \label{pilot-study}
We built Moa using Django, MySQL, JavaScript, HTML, and CSS. We used Fly.io\footnote{\url{https://fly.io/}} for deployment. For email notifications, we used SendGrid and configured it in Django's settings.py. To ensure consent boundaries work robustly, the first author created test cases and conducted pilot studies.

\subsubsection{Pilot User Studies} \label{pilot-study}
We conducted pilot user studies with five participants, who were PhD students and a postdoc who recently graduated. All participants were compensated \$20 per hour.

\paragraph{Improving the usability of consent boundary}
Participants found the consent boundary interface overwhelming because there were too many fields. Thus, instead of showing everything at once, we let users select a tab and only showed the relevant boundary dimensions (Figure \ref{boundary-dimensions}).

\paragraph{Making it clear what consent boundary features do}
We noticed that it took time for some participants to grasp the connection between their account settings and the visibility of posts and comments. For example, a woman participant did not understand that if they left their user information blank, they could not see posts from {women who used a consent boundary of restricting their content to only women.} Thus, we decided to include a short tutorial in Moa’s sign-up and account settings interfaces, which explains how account pages affect what users can see.
% Thus, we decided to add a short tutorial on consent boundaries, especially about the link between users' account and what they can see on Moa, to Moa's sign-up and account setting interfaces.

\paragraph{Emphasizing that Moa was built by a PhD student}
Some participants emphasized that knowing Moa was built by a student was important for them. We thus highlighted this on  a separate page.

\section{Phase 3: Field Study}
To evaluate Moa, we conducted a field study for almost 3 weeks with 47 PhD students in two computing-related PhD programs at a public R1 university.\footnote{One PhD program has over 100 PhD students, and the other has over 200 PhD students.} This study was exempt-approved by our institution's Institutional Review Board. 

\subsection{Methodology}
We sent recruitment emails about Moa and the field study to all PhD students in the two PhD programs. The study was also publicized on PhD-student-only Slack channels.
After participants gave their consent, successfully verified their identity, and signed up, they were sent an email that their sign-up was approved, and to start using Moa after reading a tutorial about Moa. 

We wanted participants to use Moa in a natural setting. Participants were asked to log in at least twice per day, but this was a recommendation. 
% Other than asking participants to use the system for at least a week, we did not give particular instructions. 
Participants who completed the post-study survey were compensated \$10. Those who participated in interviews were compensated \$30/hour. Among participants who posted or commented at least once, five were selected to receive \$100. 
The likelihood of winning increased based on number of posts and comments.

\subsubsection{Addressing the ``cold start'' problem} \label{addressing-cold-start}
A major challenge was attracting users to a new online space \cite{resnick2012starting}. To tackle this, the first author made 4 posts as seed content based on their experiences, and then nudged 5 early users to comment. Participants were not aware that the seed content were posted by the first author, which was an approach used by Reddit's founders~\cite{reddit2012how}. Once we recruited around 20 participants, we sent two follow-up emails about user activity and the first author's motivation for developing the system.

\subsubsection{Data collection and analysis}
After a week of using Moa, users were asked to complete a survey and invited to an interview. 31 participants filled out the survey, and we interviewed 14 of them. We also collected and analyzed log data, with users' consent.

\paragraph{Follow-up survey and interviews}
Using Likert-scale questions, we asked participants how they perceived their experience on Moa and the concept of consent boundaries. We also included one open-ended question about what, if anything, the participant wanted to change or keep about Moa. Lastly, we asked if participants were interested in a follow-up interview.

Half of the interview participants used consent boundaries (7/14; Table \ref{table:fieldstudy-interview-participants}). 9 out of 14 participants posted at least once. Of the 5 participants who did not post, 4 commented at least once. 
% One participant did not post or comment, but we included them because they expressed interest in consent boundaries in their survey response.
All interview participants gave consent to the first author to access their posts and comments. 

We asked interview participants how they decided to use, or not use, consent boundaries. For those who did not use consent boundaries but still posted or commented, we aimed to understand what enabled them to do so. The interviews lasted between 20 minutes and 45 minutes.

The first author conducted deductive and inductive coding of interview transcripts and answers to the survey's open-ended question. During the process, the first author periodically discussed the themes with the second author.

\paragraph{Log data}
We also analyzed the data of posts and comments, along with user activity. This was communicated to the participants in advance of the study, via the consent form and Moa's interface.

\subsubsection{Limitations}
Our field study's limitation is that we focused on the context of the United States.  
Future work should explore how the findings from this work could be applied to other regions' contexts. 
Another limitation is that we deployed Moa in two PhD programs. In the future, we aim to conduct a longitudinal study on a larger scale to report more robust findings.
% We plan to expand Moa to other institutions soon.

\subsection{\final{Ethics Statement}}
\final{We sought to make the terms of using Moa very clear to all users, given that our study setting is very sensitive. We indicated both on the platform and in recruiting emails that the moderator (first author) had access to all content and meta-data. No posts were excerpted verbatim from the paper, and all references to posts were paraphrased. We made it clear to users that the moderator was monitoring the platform on the backend in case harassment or trolling occurred.}

\final{We also took care in deciding methods for user disclosure and use of personal background and identity. We designed our system based on one set of best practices in consultation with several peers with diverse gender identities. No participant was forced to disclose any part of their identity. And, we allowed users to self-describe any consent boundaries as they felt were relevant in an open-ended text box (Figure \ref{boundary-dimensions}).}

\subsection{Phase 3 Findings}
% We report the findings from the field study, by combining insights from the post-study interviews, survey, and log data. 
First, we report user activity on Moa and how users perceived the platform. Then, we discuss findings on consent boundaries and on features that enabled users' participation.

\subsubsection{\textbf{Moa enabled a range of real conversations about PhD advising relationships.}} \label{overall-success}

Over 19 days, a total of 47 PhD students created accounts and logged onto Moa at least once. 
All users submitted at least some kind of user information when signing up, and most did not make changes to it.\footnote{8 users (17\%) submitted more information afterwards, and all updates seemed valid (i.e., users did not falsify information). The updates consisted of submitting advisor name, advising dynamic experiences, whether one is an international student, gender, race, year in PhD, and PhD program.}
72.3\% of the users submitted their current advisors' names and 21.3\% submitted their prior advisors' names. 36.2\% indicated that they were from PhD program A and 55.3\% indicated they were from program B. 
We report participant demographics in Appendix.

\begin{figure*}[t!]
    \includegraphics[width=0.32\linewidth]{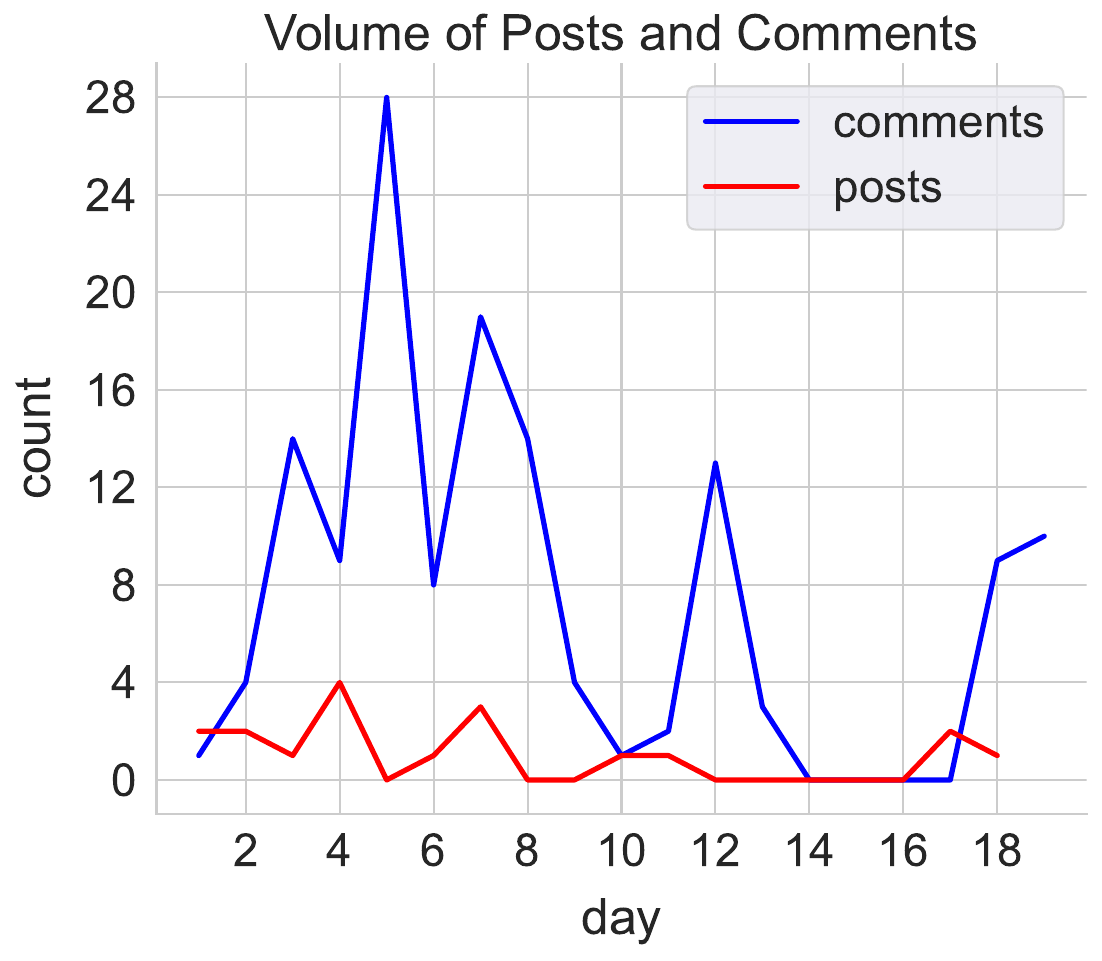}
     \includegraphics[width=0.32\linewidth]{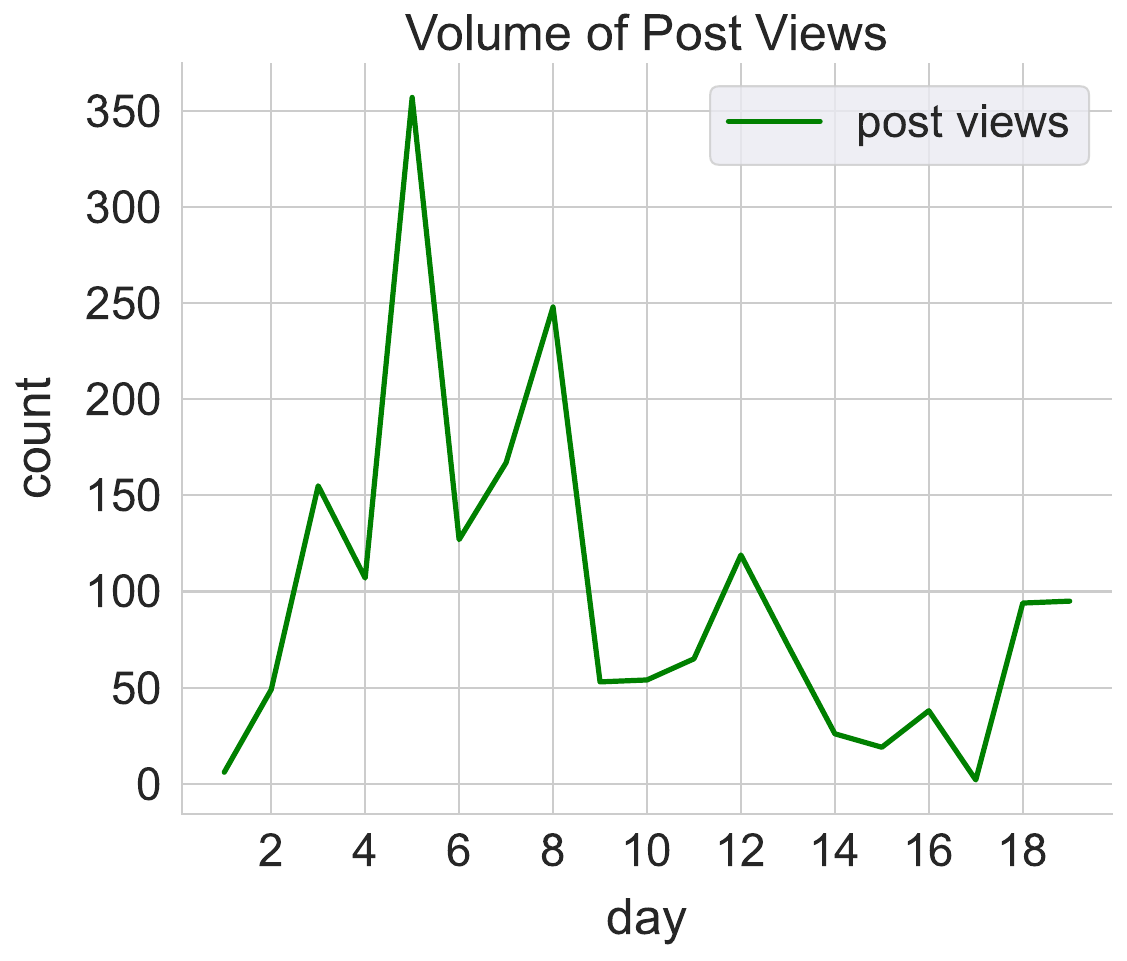}
 \includegraphics[width=0.32\linewidth]{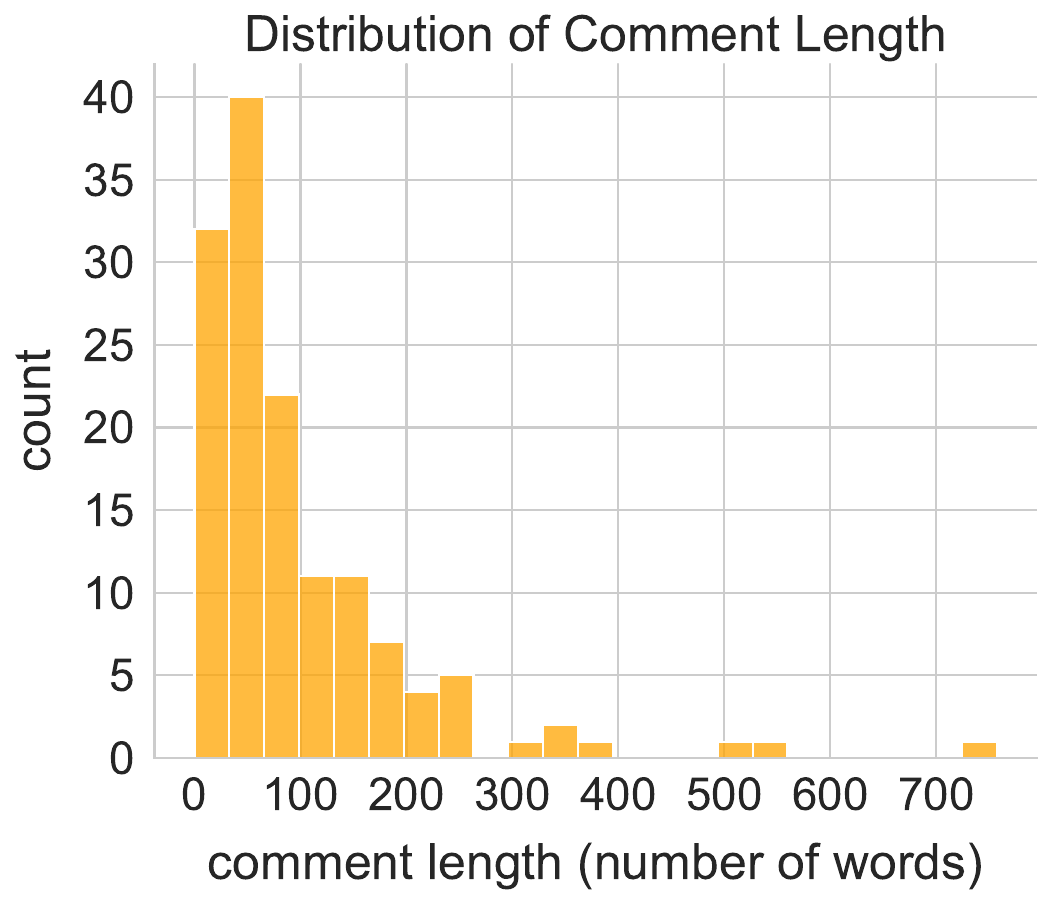}
    \caption{The first graph shows Moa's daily post and comment volume, the second shows daily post views, and the last shows comment length distribution. On average, posts had 7.7 comments, with a maximum of 17. The daily post views averaged 98.2. Comments averaged 98.5 words, with the longest at 757 words.}
    \label{fig:moa-activity}
\end{figure*}

15 users created 18 posts and 31 users left a total of 139 comments\footnote{These numbers exclude the seed content the first author created.} and users viewed posts on Moa 1,866 times (Figure \ref{fig:moa-activity}). Moa's users overall engaged with each other. On average, posts had 7.7 comments, and at most, one post received 17 comments. The comments had 98.5 words on average (Figure \ref{fig:moa-activity}).
Posts that asked for advice about their advising situation especially received comments with detailed advice---the longest was 757 words.
Users tended to thank each other for the advice.

\begin{figure*}[t!]
    \centering
     \includegraphics[width=.75\linewidth]{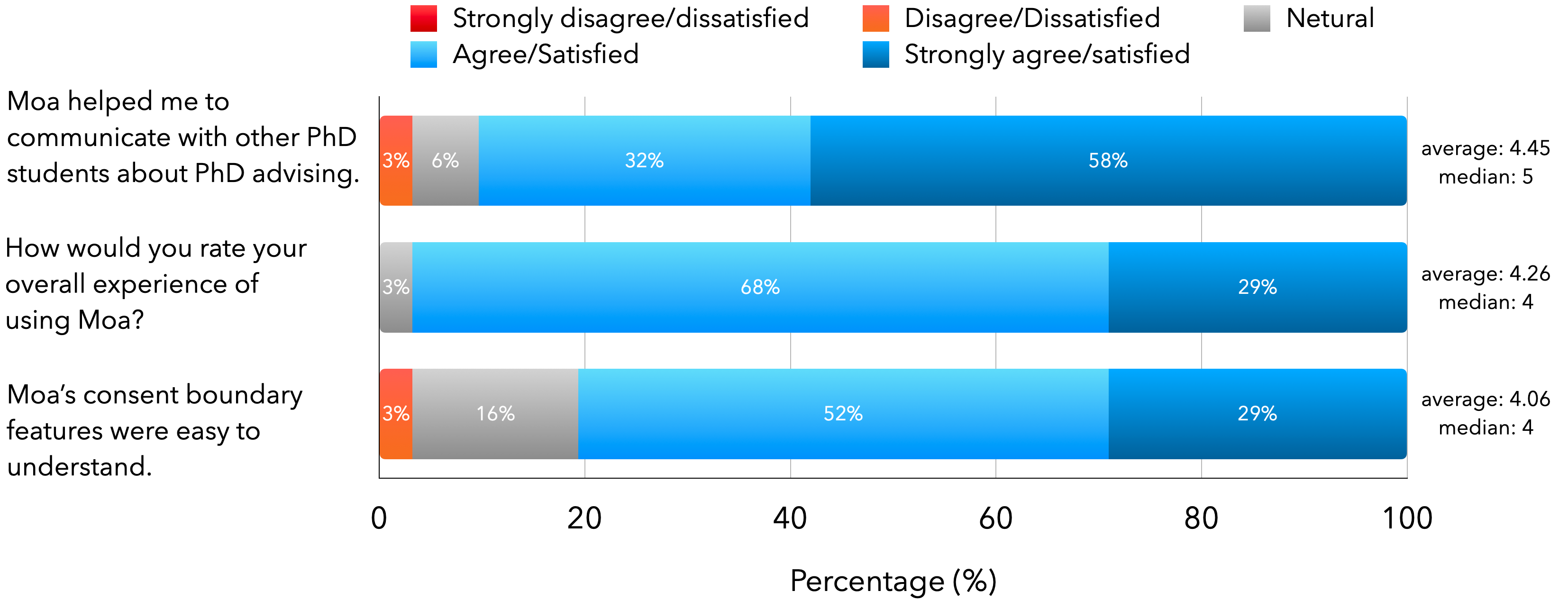}
    % \vspace{10px}
\caption{Results of the post-study survey (1=Strongly disagree/dissatisfied; 5=Strongly agree/satisfied). Participants strongly agreed that Moa made it easier to have conversations around PhD advising (average=4.45; median=5) and were satisfied with the overall experience (average=4.26; median=4). Many found consent boundaries easy to understand (average=4.06; median=4).}
\label{survey-results}
\end{figure*}

% \begin{quote}
% ``\textit{I really hope Moa can continue operating. Just like csranking.org or something. [... ] Then let's try to find funding!}''-P2
% \end{quote}

% \begin{quote}
% ``\textit{I think Moa is really helpful because I wanted to hear about, read about, just learn about other challenges that people have had in their advising experiences because it's something that's difficult to talk about. And I wanted to learn about what was the spectrum of experiences people were having, to sort of situate mine.}''-P1
% \end{quote}

% \begin{table}[]
% \begin{tabular}{lr}
%                                     & \multicolumn{1}{l}{\textbf{ratio}} \\
% \textbf{women}                      & 48.9\%                             \\
% \textbf{men}                        & 31.9\%                             \\
% \textbf{non-binary}                 & 2.1\%                              \\
% \textbf{did not submit information} & 17\%                               \\
% \textbf{}                           &                                   
% \end{tabular}
% \end{table}

With respect to the kinds of interactions that occurred, among the 18 original posts, 9 asked for advice (see Table \ref{post-categories}).\footnote{Four posts were deleted; we did not include these in the final dataset.} These included a range of sensitive posts where users asked for advice. The most sensitive posts were about seeking guidance on whether to switch advisors or add a co-advisor, as well as dealing with mental health struggles after switching advisors due to harm from a prior advisor. Other topics also included navigating gender dynamics with another advisee, seeking to diagnose reasons behind communication issues, how to cope when advisor is not receptive to feedback, and how to better work with their advisor given their mentoring style. In three posts, some users proactively shared personal experiences, such as how they switched advisors. 

Overall, users' reactions showed that they received real benefit
from Moa. {For example, P14 reached out to the first author on Slack saying ``\textit{Thank you for making this system. I was in deep need for it.}''}
In interviews, participants often commented that Moa enabled them to learn what good advising looks like and what ``is not normal.'' Survey results (Figure \ref{survey-results}) also showed that many participants strongly agreed that Moa made it easier to have conversations around PhD advising relationships (average=4.45; median=5).  

\begin{table}[t!]
\small
\begin{tabular}{lc}
                                                    & \multicolumn{1}{l}{\textbf{\# of posts}} \\\midrule
\textbf{Asking for advice about PhD advising relationships} & 9 (50\%)                                           \\
\textbf{Sharing prior experiences}           & 3 (16.7\%)                                          \\
\textbf{Sharing positive stories}            & 2  (11.1\%)                                          \\
\textbf{Sharing high-level opinion about PhD advising}                     & 1  (5.6\%)                                          \\
\textbf{Discussing PhD-related topics in general}                                                 & 3  (16.7\%)  
\end{tabular}
\vspace{7px}
\caption{Number of posts per topic. Out of 18 posts, half were seeking advice on PhD advising relationships.}
\label{post-categories}
\end{table}

\begin{table*}[t!]
\small
\begin{tabular}{c|ccc}
\multicolumn{1}{l}{}  & \textbf{Activity on Moa} & \textbf{Consent Boundaries Used} & \textbf{Set Default Consent Boundary} \\\midrule
\textbf{P1}    & posted/commented     & advising challenge  & no                                      \\
\textbf{P2}   & posted/commented       & none      &    -                                       \\
\textbf{P3}      & commented       & none             &  -                              \\
\textbf{P4}      & posted/commented      & none     &    -                             \\
\textbf{P5}   & posted/commented        & advising challenge       & no                                 \\
\textbf{P6}     & commented        & advising challenge, advising status, race & yes + customized per comment                                            \\
\textbf{P7}     & posted/commented           & PhD program        & no                         \\
\textbf{P8}          &   none                              & none   &     -                 \\
\textbf{P9}       & posted/commented        & none                 &  -                 \\
\textbf{P10}    & posted/commented       & none              &         -              \\
\textbf{P11}      & commented          & gender       & yes                                  \\
\textbf{P12}     & commented                    & advising challenge     & no          \\                
\textbf{P13}      & posted/commented     & gender             & no                   \\
\textbf{P14}     & posted/commented        & none       &       -                
\end{tabular} 
\vspace{5px}
\caption{Overview of users who participated in the field study, post-study survey, and interviews. The first column shows whether each participant posted or/and commented. The second column shows whether/how each participant used the consent boundary feature. For those who used consent boundaries at least once, the third column shows whether a participant has set their default, account-level consent boundary. ``yes'' means the user has set at least one boundary to a non-default value. ``no'' means everything was at default value. We put ``-'' for users who did not end up using consent boundaries.}
\label{table:fieldstudy-interview-participants}
\end{table*}

% For example, one poster emphasized they were risk-averse, but still posted on Moa because it provided something that social media like Reddit did not. When asked what it was, the participant said:
% \begin{quote}
%     {``\textit{Some people explicitly said like, `Thank you,' to me. That really made me happy and incentivized me to write more.}'' -P7}
% \end{quote}

% {Many participants commented that Moa provided trustworthy and high-quality advice, which partly was due to knowing that everyone was from the same institution. At the same time, being in close proximity meant that users had to navigate the heightened risks of revealing one's identity.}

% \paragraph{Users who used them}

% For users who did not use them, but saw at least one boundary, they perceived ...

% \subsubsection{Users tended to submit at least some kind of information}
% account settings

\subsubsection{\textbf{Consent boundaries were used by about a quarter of posters/commenters \final{(N=7/31)}, who tended to set them per post or comment.}} \label{consent-boundary-usage}

7 out of 31 users who posted or/and commented (22.6\%) set consent boundaries for at least one post or comment (Table \ref{table:fieldstudy-interview-participants}). 3 out of 7 users used it for posting and 4 used it when commenting. For the three posts, they were sensitive in nature, although the degree of sensitiveness differed---one was about their advisor not being receptive to feedback, another was about gender dynamics in their advisor's group, and the last one shared their experience of switching advisors. In the comments, users asked questions, shared one's experience (e.g., having difficult conversations with one's prior advisor), and expressed agreement or sympathy.

Contrary to what we expected, almost all users set a \textit{distinct} consent boundary \textit{per comment or post}. The majority (5/7) did not touch their account-level consent boundary at all, and only P11 used their account boundary setting as it is (Table \ref{table:fieldstudy-interview-participants}). Users mentioned they found the consent boundary being ``right there'' and found it easy to apply them for each comment. More interestingly, some interview participants compared consent boundaries against existing privacy settings, which they noted were binary (public versus private). In contrast, participants thought consent boundaries are more contextual and thus wanted to apply them per post or comment.
% For example, P1 described existing platforms' privacy settings as a ``blanket setting over everything'' and being binary, while consent boundaries were related to ``an interaction level.'

% \begin{quote}
%     ``\textit{Having all of the options available to me and then refining it from there just felt easiest in the moment, instead of having to remember, `Oh, if I say this, only people who had this boundary setting will be able to see what I say for any post that I make', whereas taking it on a post-by-post basis felt just more manageable.}'' -P1
% \end{quote}

\begin{quote}
    ``\textit{I kind of felt like I have so many different types of experiences that are for different groups that I don't want to put something and then have to go back and undo it. So I felt like per post was nice. That way I can choose who sees what when I make the post...}''-P13
\end{quote}

The survey results showed that most users found the consent boundary features easy to understand (Figure \ref{survey-results}). 
% As shown in Figure \ref{survey-results}, the majority of users agreed with the statement ``Moa’s consent boundary features were easy to understand'' (average=4.06; median=4).
However, three out of the 31 survey participants commented that they found consent boundaries difficult to understand. 
% The interviews also showed that a few participants did not fully grasp how their account information and account-level consent boundary are different, or linked. 
Some interview participants also wished the feature was less overwhelming, although they understood why we included each dimension.
% \begin{quote}
%     ``\textit{I liked that there were a lot of options, but I think also when I was making a post or thinking about when to restrict things, I was like, this is so many options. }''-P13
% \end{quote}

\paragraph{\textbf{Restricting boundaries}} The feature for restricting consent boundaries was not used frequently, although one participant used it twice. P6's comment was public at first, but they later made it visible to users who are Asian. Then, they updated the boundary so that the comment is visible to those who are Asian \textit{and} have switched advisors.

\subsubsection{\textbf{Consent boundaries were used for not just defensive boundary-setting but also proactive audience curation}}\label{consent-boundary-usage-reasons}
% The survey results showed that those who used, or attempted to use, the consent boundary features, tended to agree it helped in feeling more comfortable in engaging in discussions on Moa (average=3.67; median=4). 

The interviews showed that users tended to use consent boundaries differently.
For some, consent boundaries were used to filter out specific kinds of users for privacy; for others, they served as a mechanism to proactively curate an audience for various reasons. 

Specifically, three participants (3/7) used consent boundaries to protect their privacy by not making their post public. These participants posted or commented to ask or share personal experiences about their advising relationship, but did not want to increase the chances of revealing their identity---not just because of their privacy, but also because of their advisor's. For example, P5, who used boundaries about certain kinds of PhD advising dynamics when posting a question, said:
% One participant (P5) used boundaries related to certain kinds of PhD advising dynamics and experiences.

\begin{quote}
    ``\textit{I wasn't sure if I wanted everybody on on the platform to be able to see the post, mostly because it is kind of like a personal issue. And then I was also worried that people could identify me from it, and honestly, I think I didn't want it to reflect badly on my advisor.}''-P5
    % […] So I think that's why I decided to apply a consent boundary.
\end{quote}

Interestingly, the rest of the interview participants who used consent boundaries viewed it as a way to \textit{proactively reach} certain groups of people, rather than preventing some people from viewing their post. Their reasons included wanting to reach people who would better understand them, provide accurate responses, or find the content helpful. For example, P1 used boundaries related to certain kinds of advising challenges. To them, privacy concerns were not a reason for using consent boundaries, because they  refrained from writing sensitive content. But, P1 still wanted to interact with people who could provide the most accurate insights.

\begin{quote}
    ``\textit{ So I think I mostly used it [consent boundaries] to try to make sure that my comment reached the people that I wanted it to reach as opposed to avoiding people I didn't want it to reach, if that makes sense.}''-P1
\end{quote}

In particular, two consent boundary setters emphasized wanting to interact with those who could understand them. For example, P6 wanted to help other PhD students by sharing their experiences, but in a way that minimizes chances of getting hurt by reactions from people who cannot understand them. This participant was the most active in using different kinds of boundaries. P6 explicitly said it was due to the desire to have an audience who could understand them, rather than reducing privacy risks.

\begin{quote}
    ``\textit{I think privacy in this specific experience might be less relevant because I feel like whatever consent boundary I set, I can be identified by sharing my experiences. [...] Or another way to put this is that I don't expect or really want to be understood by all people.}''-P6
\end{quote}

\subsubsection{\textbf{{Some participants found value in showing consent boundaries, while others chose to not make them visible.}}}
% Most users did not show their consent boundaries, even for public content. Only 6 of 18 posts had visible boundaries, with 2 of 6 being non-public (i.e., restricted by consent boundaries). Among 139 comments, only 26 showed boundaries, and 4 of these were non-public.
Users showed mixed behaviors when deciding whether to display consent boundaries on posts or comments. Of the seven users who used consent boundaries, four displayed them, two kept them entirely invisible, and one alternated between showing and hiding their boundary.

% {One participant who did not visualize their consent boundaries perceived them as private and sensitive information, even when their post was public, and intentionally decided to hide the information.}
% \begin{quote}
%     ``\textit{I saw the consent boundary as more of a private setting [...] and that's why I thought that I probably want to keep it as hidden and not visible to others because I don't want people to know how I am, what's my behavior of sharing posts or what audiences am I selecting into sharing those posts.}''-P4
% \end{quote}

Those who used \textit{and} showed consent boundaries thought there was value in signaling whether a post was a ``closed thread'' for specific kinds of users. They noted that without visible boundaries, others might assume the post is public or be unsure about its visibility, making them hesitant to respond. For example, P5, used and showed boundaries of limiting their post to those who have experienced specific advising challenges, thinking that it would ease users to respond, since their comments will be viewed by those who had the same experience.
\begin{quote}
    {``\textit{I sort of just wanted to give some transparency to anyone who felt like they want to respond to this with a personal story. [...] I felt like if I were to set the standard in the first post, then maybe people would be more comfortable knowing that it had a limited audience to begin with.}'' -P5}
\end{quote}

In contrast, some tended to think that once the intended audience sees the post or comment, then the consent boundary itself is unnecessary information. That is, they used consent boundaries for audience selection, rather than for showing bits of their identity. P6, who was the most active in using consent boundaries, said:
\begin{quote}
    {``\textit{I feel like in reality, when a person sees a post or a reply, whatever the consent boundary is visible to them, doesn't change the fact that they can see this reply. So I think [shown consent boundaries] is unnecessary information from that point.}'' -P6}
\end{quote}

Finally, a few interview participants did not want to show their consent boundaries because they thought it was a private information they wanted to keep to themselves.

% {One other participant (1/3) showed their consent boundary about experiencing a particular advising dynamic so that other users could quickly find the comment, using the boundary's keyword.}
% \begin{quote}
%     {``\textit{...somehow I felt like this [boundary] keyword has some relevance when a person is seeking help for a particular topic. For example, in Reddit, what we do normally [is] we are not just browsing the post and find something that is relevant to me. In most of the cases I'll use the search engine because I'm searching for a particular topic.}''-P12}
% \end{quote}

% \begin{quote}
%     \textit{There were actually cases where I saw a visible consent boundary in some comments, and I think it was a little surprising, even though I took that strategy myself. But yeah, I think it was surprising just because not a lot of people were sort of using that feature to make it visible.} -P5
% \end{quote}

\paragraph{\textbf{{Some participants thought seeing consent boundaries was helpful.}}}
Some interview participants thought seeing consent boundaries was helpful, because it provides more context on why the poster is reaching out to a specific group of users. For example, P11 noted that seeing a a women-only boundary on a post about gender dynamics made them want to engage with it.
% \begin{quote}
%%%     {``\textit{I don't see it [consent boundary visualization] being much important for someone reading. If I'm going to read it, I'm going to see the point of view to understand what the writer is writing. I don't see much the need seeing which group I am in.}''-P3}
% \end{quote}
% At the same time, for P8, the visibility of consent boundary triggered privacy concerns---this was based on an inaccurate mental model of consent boundaries, as the participant did not understand that the comment's consent boundary was currently visible to only those who can see the comment.
\begin{quote}
    {``\textit{Like since they are showing a woman-only boundary, I will be very likely to respond to this post in a very, very honest way and support this woman student.}''-P11}
\end{quote}

{A few participants also raised ideas on making it possible for users to share parts of their identities in comments, regardless of setting a particular audience. For example, P13 said that this can be useful when a user finds a post that resonates with them because it is about a similar experience they have had, and wants to signal the connection to the poster.}
% \begin{quote}
%     {``\textit{I think maybe it could be interesting to have, if you're going to reply to something that you do want to know that they've had similar experiences, you could mark who you are within the amount that you want to, but without restricting it [audience].}''-P13}
% \end{quote}

% \paragraph{One user applied the consent boundary rule to the original poster}
% By default, when a user replies to a post or a comment, it is visible to the poster/commenter (see Section 5.2). However, a user can toggle the setting so that their boundary is applied even to the poster. This setting was used once, when the poster asked a user about their experience of bringing up a difficult topic with their prior advisor. The commenter replied with a lengthy comment, but applied two consent boundary rules (one related to racial identity, and one related to whether the user has switched advisors), while forcing the rules to the poster as well. This means that the poster was not able to read the comment if they did not meet the two criteria. We suspect that this is likely because they perceived the answer could be too revealing of their identity.

\subsubsection{\textbf{Reasons for not using consent boundaries}}  \label{not-using-boundaries}
Most posters and commenters did not ultimately set consent boundaries, and they also cited several reasons for not doing so.
% \paragraph{Fundamental mistrust}

\paragraph{\textbf{Wanting to have a broad audience}}
One major reason was that they wanted to enable their post to reach the broadest audience, and thus made their content public. For example, P4, who publicly wrote about advice regarding advising relationships, said they wanted it to reach a lot of people. This was also the case for three interview participants who asked questions about the advising challenges they faced. 
These participants, however, were not posting publicly out of comfort. They disguised their content---e.g., by altering the gender of their advisor---so that readers could not identify them, especially given that they are from the same institution or program. 
\begin{quote}
    ``\textit{I intended that (i.e., set their post to public) because I want to get advice from everyone who can be related and I don't know who they are.[...] Moa’s users are highly related to me, so they can give me more direct and valuable advice. But at the same time, it makes me conceal myself more because I feel like they can easily guess it’s me.}'' -P10
\end{quote}

% They said they were able to post publicly because they only shared their advising situation to one person. They highly trust that person did not tell others in the department, so because of this, they felt confident that no one could realize it was them. 

\paragraph{\textbf{Anonymity and PhD student verification were enough}}
For some users, anonymity through usernames, combined with the strict identity verification imposed by the system, were sufficient. 

\begin{quote}
    ``\textit{For me personally, I'm an international student and I am kind of in fear of authority, but at the same time, I kind of trust you. Everyone on here is a student, so I don't mind if everyone is a student, I don't think they will do anything.''}-P3
    % ``\textit{I think with that one, if my name was on it, I would've used consent boundaries. [...] I don't know, I would've restricted it somehow, but I think since my name wasn't on it, and it was not a very sensitive type of question, it was just kind of like, `Hey, I'm kind of new here and I don't really know what I'm doing. What do you guys do?' It kind of felt like a safe one to make public. [...] But like I said, if I think of my name was associated with it, I wouldn't want to ask it as public.}''-P13
\end{quote} 

The multiple username feature was popular---20 users (64.5\% of posters/commenters) created 54 distinct usernames, aside from their default username, when posting or commenting (average=2.7). 23.4\% of all users changed their default usernames at least once, as well. The interviews and survey showed that this feature helped users feel reassured that others would not learn their identity.

% \begin{quote}
%     ``\textit{I actually really liked the way i could post things under different usernames — that way I felt it was less likely that other people would be able to piece together my identity; correspondingly, I’d change my style of writing along with the username to make it harder for others to identify me.}''-P9
% \end{quote}

% \paragraph{Could always restrict a consent boundary later}
% One participant said that because they noticed they could always restrict their consent boundary later, they started by leaving public comments.

\paragraph{\textbf{System creators' and other users' credibility also helped}}

Even for participants who did not know the first author well, Moa being developed by researchers in the same institution and detailed explanations of how we treat users' consent increased Moa's credibility. The first author's revelation (in a recruiting email) of having switched advisors also helped. For example, P13 said:

\begin{quote}
    ``\textit{I think one was how much you were respecting our anonymity and our consent boundaries and things like that. I kind of felt it empowered me to post about things that maybe I wouldn't have otherwise. [...] The description was part of it, where it was like, `We're making this platform for this purpose, and it's made by people who have had experiences that they've wanted to talk to other people about.'} ''-P13
\end{quote}

% \begin{quote}
%     ``\textit{And I know because in [institution name] privacy and everything is important, and you said [on Moa's webpage and consent form], I'm going to save these but remove these and so on. All these descriptions made me like, okay, I trust the person, although I don't know them but I trust them, since they're within the institution where I'm at.''}-P14
% \end{quote}

Users' credibility also incentivized some users to post. Interview participants said they could get more trustworthy advice because they knew the users were from the same institution, and not just a random person on the internet.
For example, when asked what enabled them to post, P14, whose post asked for advice about navigating advising challenges,  said:
\begin{quote}
    ``\textit{I think several factors. One is, well, the first thing I can seek advice. The second thing I was looking for something where I am anonymous. On social media and everything you're mostly known, or if I go to my lab mates or anything they know who I am. The third thing is I know it's within [institution name], so I'm sure where they are but I'm not sure who they are. It's like I'm sure the people who reply to me are actually PhD students.}''-P14
\end{quote}

\paragraph{\textbf{Insufficient trust in system}}
Some participants refrained from writing sensitive content (and did not use consent boundaries) because they needed more time to build trust in the system. % who tried using the consent boundary feature but did not end up using it, lack of trust seemed to be one reason. While these participants agreed that the consent boundary features is helpful for posting very sensitive content (e.g., more egregious harms from one's advisor), they emphasized they still need to develop more trust in Moa to post more sensitive information in the first place (and thus use the consent boundary feature). 

\subsubsection{\textbf{Signs of users wanting closer interactions with each other}}
A few interview participants mentioned the possibility of directly reaching out to other users on Moa or wanting to meet in-person. That users wanted closer interactions was surprising to us, because it contradicted what the majority of our Phase 1 participants said.\footnote{Many Phase 1 participants did not want conversation on Moa to be too private and some also emphasized that they would never reveal their identity (Section \ref{shallow-interactions}).} For example, a few Phase 3 interview participants said they wished Moa had a chat interface so they could more easily message others, while another said there was one particular user who wrote a post that made them want to connect in-person. These show that through Moa, some users developed trust in others, to the extent of wanting to take the risk to further interact with them.
\begin{quote}
    ``\textit{I guess if I know that only people with that consent boundary are now seeing the post and they replied, after the initial post, I would rather have a conversation with them. [...] I believe another feature that could help for very sensitive situations where I'm really not comfortable sharing it as a post would be having some kind of a private messaging.}'' -P4
\end{quote}

When asked why, participants mentioned wanting to get more advice because they desperately needed to resolve their advising situation, and to obtain accurate signals about a person \cite{donath2007signals}.
\begin{quote}
    ``\textit{There is a really impactful post [on Moa]. If I see an impactful post [like that], then I really want to reach out and get advice from them. [...] I want to meet in-person, and talk sincerely because their post is very sincere, and I'm touched by that.}'' -P10
    % ``\textit{Once there were a bunch of conversations and then there was one that I thought really did resonate with me that I wanted to learn more about, even if it was a very problematic topic, if there was like, okay, based on all of these topics that we've discussed, there's going to be these in-person advising anonymous meetings or something based on these conversations people have said, and we'll sit there and have a safe space of being able to talk about this in a non-technology facilitated way, but using this as a way to filter what those topics would be, I think that's something that I would probably use more.[...] I sort of use them [Moa's consent boundaries] as a filtering criteria where I can set who I want ... People who have similar concerns to me, I know that my content is going to reach them. Being able to use that as a way to get into the same space, even if it's a temporary space with people with similar concerns. }''-P1
\end{quote}

{It seems that while some risk-averse participants considered Moa to be useful, they believed that giving or getting more sensitive information should happen outside the platform, with discussions on Moa being a basis for such connections. For example, P1 pointed out that Moa could be a starting point for in-person meet-ups.}

\begin{quote}
    {``\textit{If there was like, okay, based on all of these topics that we've discussed, there's going to be these in-person meetings based on these conversations, and we'll sit there and have a safe space of being able to talk in a non-technology facilitated way, but using this [Moa] as a way to filter what those topics would be, that's something I would probably use more.}''-P1}
\end{quote}

\subsubsection{\textbf{Ways to improve Moa}} We describe the most interesting feedback on improving Moa.

\paragraph{\textbf{Using Moa to help faculty learn about mentoring}} 
One interview participant said while they were not so sure how it should happen, they thought it would be worthwhile to invite faculty to Moa, as they could gain exposure to students' experiences and learn about mentoring.

\paragraph{\textbf{Expanding the user pool while providing  audience control via consent boundaries}}
Some interview participants wished there were more users, and not just from two PhD programs. However, even these participants perceived benefits of interacting only with those from the same program, and said the consent boundary could be a way to easily show their post to them.

\paragraph{\textbf{Making consent boundaries more usable}}
Multiple interview participants wished the consent boundary interface was more usable. One participant suggested a feature that gives users a sense of the audience size when applying a consent boundary. Related to the flexibility of applying different consent boundaries, one participant recommended using a computational modeling approach so that the system could recommend consent boundary settings based on each post.

% \begin{quote}
%     \textit{``I suggest that Moa could implement a predictive feature that highlights and reminds users when they might want to adjust the consent boundary based on the specific content of their post (and hide it when it is not a consent sensitive post?).''}-P11
% \end{quote}

% \begin{quote}
%     ``\textit{It would be nice to get an estimate of how many people can see the post (as if I'm seeking help, I wouldn't want the boundary to be so restrictive that only a few people can offer help).}''-P5
% \end{quote}

\subsubsection{{Summary of Results}}
{Moa enabled a range of sensitive discussions, with a quarter of participants using consent boundaries---typically a separate boundary per post or comment. Users tended to use consent boundaries differently---some used them to simply remove users to reduce privacy risks; others used them for proactive audience curation, to reach those who are sympathetic, could provide useful input, or find the content helpful. Some participants valued displaying their consent boundaries, while others chose not to, because they used them solely for audience configuration. Finally, a few participants expressed interest in directly contacting another user on Moa or meeting them in person, which was surprising compared to the results from Phase 1. }

% \begin{quote}
%     ``\textit{I think you can broaden users. You said that you only invite the PhDs in [names of the two PhD programs]. I think we all know each other and all [in their PhD program] are connected. I think they're likely to guess it's me because my labmate mentioned it [their advisor's style] to other people.}''-P10
% \end{quote}

% \paragraph{\textbf{Adding trigger warnings}}
% Some participants wanted a feature inverse of consent boundaries---something similar to trigger warnings, given that the topic is related to interpersonal harm. 
% \begin{quote}
%     ``\textit{Just like in Facebook, I might want to change, "I don't want to see these kinds of posts." I might want to do that for Moa because I believe a lot of content is also sensitive in nature.}''-P4
% \end{quote}

\section{Discussion and Future Work}
% Here, we reflect on our results mainly on two levels: synthesizing our findings into something of a ``recipe'' for enabling sensitive discussions online; and discussing the novelty and value of consent boundaries. 
% Then, we reflect on how a consent-centered approach differs from privacy-centered approaches in system design. Finally, we consider how systems like Moa could be further adapted for wider-scale use and collective action.

\subsection{Recipe for Enabling Sensitive Conversations and Ally Discovery Online}

Combined, the existing literature on enabling sensitive conversations makes many recommendations, but they are scattered across multiple papers ~\cite{ma2016anonymity,abdulgalimov2020designing,leavitt2015throwaway}. Here, we pull them together to provide a state-of-the-art, socio-technical ``recipe'' for such systems. We present it in two layers: One layer mentions the high-level objectives sought (indicated by the boldface headings below); the second contains specific features that enable elements of the first layer (boldface, in parentheses).

\paragraph{\textbf{Anonymity (pseudonyms, multiple usernames, no profile pages)}} As prior literature shows \cite{ma2016anonymity,abdulgalimov2020designing}, and our Phase 1 study confirmed, providing users anonymity is important. Complete anonymity, however, can make discussion threads difficult to follow. A good compromise is to allow users to create pseudonymous usernames---many of Moa's users created multiple pseudonyms.
Avoiding profile pages, as Whisper does, also enhances anonymity. None of our field study participants said they wanted a profile page, and one participant explicitly said they liked not having them. %It meant users could not easily keep track of what others are posting or commenting, as one participant explicitly pointed out.

\paragraph{\textbf{Credibility of system (baseline system security, sympathetic sponsors, consent emphasis)}} System credibility requires technical privacy and security protections as a baseline, and Moa assured users of that, but there is also a social element. Confirming findings from online systems in general~\cite{lothian2021archive}, Phase 1 study showed that they would use Moa more if it were run by people they trusted, especially a PhD student peer. A few participants of Phase 3 also noted that Moa's emphasis on consent was intriguing to them, and enhanced the system's credibility.

\paragraph{\textbf{Credibility of users (strict identity verification, affiliated user base, moderation)}} Credibility of users is also critical, as prior work has noted \cite{abdulgalimov2020designing}. One important factor is for the user base to be limited to some sort of affinity group. In Moa's case, all users were PhD students at the same university. Phase 3 participants noted that one reason they found Moa more useful than, say, Reddit, was because they could expect more relevant responses from peers who shared a lot of context. We also believe that it was easier to trust other users because it was a limited, familiar, and moderated community.
{In the future, we anticipate Moa could be opened to various institutions, but users would still be able to post to others within the same institution using consent boundaries.}
% Future work could investigate the degree to which size and similarity affect meaningful engagement. 
% It is not clear with Moa, for example, whether and how engagement would change if the system were opened up to all departments at the same university, or similar departments at multiple universities. 

%it is possible to keep one's anonymity by using throwaway accounts, but because the user base is so large, it was harder for participants to expect high-quality answers. Knowing the source of Moa's answers---because we strictly verified all users are PhD students in two PhD programs---made them more credible.

\paragraph{\textbf{Engaged, empathetic atmosphere (seed content, community rules, in-group moderation)}} An engaged atmosphere is essential for any thriving online community~\cite{resnick2012starting}, but for sensitive discussions, the culture must additionally be empathetic. To enable this, we believe appropriate seed content, community rules, and moderation are critical, especially at the beginning. Because many Phase 1 participants said they would lurk first before deciding they could trust the system, the first author seeded posts based on their personal experience. Two Phase 3 interview participants noted that one of the seed posts was memorable.
Furthermore, Moa's community rules emphasized Moa being an supportive community for PhD students. And, we believe in-group moderation is essential. Moa's users seemed comfortable with moderation, likely because Moa's moderator was a member of the user base. This is in contrast with prior work, in which participants raised questions about moderation, because an external research team handled it for the study~\cite{abdulgalimov2020designing}.
%probably because the moderator was a member of the user base.  a PhD student themself, while in Abdulgalimov et al.'s work, the moderators were the research team and not staff members, who were the system's major users.\footnote{The authors wrote that their system was deployed to an academic department where there were 600 staff members and over 100 PhD students \cite{abdulgalimov2020designing}.} Another reason could be that PhD student users were sympathetic to the first author to some degree, as a fellow PhD student. Our Phase 1 participants noted that while they wanted PhD students to be moderators, it meant a lot of emotional work, so they were not so sure how feasible it was.

\paragraph{\textbf{Granular control over post visibility (consent boundaries)}} For users who want to discuss topics or share experiences that are highly sensitive, granular control over who can see their posts seems to increase chances of engagement. For Moa, this was enabled by consent boundaries. The Phase 3 evaluation showed that consent boundaries lowered the barrier to posting for some users, especially for risk-averse ones. Even for users who had less privacy concerns, consent boundaries still provided benefits of reaching out to the right audience. \final{We also note that there was arguably an implicit consent boundary of all users being PhD students from the same institution. Furthermore, the users also consented to the moderator seeing their posts, which could be seen as another implicit consent boundary due to the study design. These could explain why the consent boundary's usage rate was 22.6\%---the default, implicit consent boundaries already provided an audience that made most users comfortable enough to post.}
% However, we note that consent boundaries were used only by a quarter of posters/commenters, in part due to the complexity of setting them. Consent boundaries incur a tradeoff with usability because of their complexity, and low usability can be a barrier to engagement. Future work could explore better ways to manage that tradeoff.  

\paragraph{\textbf{Feature visibility (transparent explanations, repeat feature reminders)}} Finally, it is important for all of the above to be visible to users. Messaging on the system should highlight key system features repeatedly. In Moa's case, the goals and features of the platform were noted on the landing page, emphasized in recruitment emails, and scattered throughout the platform's interface.

% \subsection{Developing systems that enable conversations that were hard to have before}
\subsection{Consent Boundaries: Values and Challenges}%{esigning Nuanced and Flexible Consent Boundary Interfaces}
While many of Moa's features are drawn from known social media designs, consent boundaries are a novel design contribution. 
Two Phase 3 participants (P11 and P13) mentioned explicitly that they had never seen anything like consent boundaries. 
Below, we discuss the values and novelty of consent boundaries, and then outline how to further improve them.

\subsubsection{Novelty and value of consent boundaries}

First, as participants noted, consent boundaries offer \textit{highly relevant, context-focused} dimensions for audience selection. In the case of PhD students discussing advising challenges, the  dimensions were related to social identity, academic experience, granular information about the institution, etc.---such dimensions are not offered by existing platform structures \cite{zhang2024form} or access control systems, most of which offer simple subsets such as ``private / friends / public'' or technically defined groups such as ``administrator, editor, viewer.''

Second and related to the point above is that Moa's consent boundaries offer \textit{finer-grained} dimensions than existing social media or access-control systems. Based on Phase 1's findings, the types of boundary dimensions span the full range of unions and intersections of various groups. And, Phase 3 participants who used consent boundaries leveraged different kinds of dimensions.

Third, our design of the implementation of consent boundaries carefully attends to the process of setting them. In contrast, most platforms treat audience selection as a rarely adjusted element that requires users to navigate complex ``Settings'' features. Our implementation honors the principles of affirmative consent~\cite{im2021yes}, by ensuring that \textit{every step} of consent-boundary-setting respects user consent. For instance, to ensure consent \emph{specificity}, users can select a boundary for each post and comment; and indeed, Phase 3 participants tended to do so.

\subsubsection{Improving consent boundaries}
Despite the potential value of consent boundaries, there is a tradeoff with usability. 
With advancements in AI, one could imagine a system could learn how a user uses consent boundaries, and then make suggestions \cite{im2021yes}. Or, a user could type their consent-granting preferences in natural language, and the system automatically suggests boundary dimensions. (Of course, such an approach might have to wait until AI models are perfect with respect to interpreting user requests in order to adhere to the principles of affirmative consent.)

Another challenge is that for some consent boundaries, whether a  user is inside of them may not be readily verifiable. For example, users can ask to have a post be visible only to others who have experienced bullying by their advisors, but whether a user has had such experiences cannot be easily confirmed. It is therefore possible that a user reports their experiences inaccurately and gains access to discussions. Strict fidelity to affirmative consent would not allow such leakage; maybe such dimensions should not be implemented. It is also possible, however, that compromises should be made to facilitate the desired communication. Future work should explore these tradeoffs.

\subsection{\final{How a Consentful Approach Differs from a Privacy-Centered Approach}}
Taking a consent-centered approach was essential to developing Moa. \final{While consent and privacy are closely related, there is a subtle difference in emphasis. Speaking strictly from an engineering standpoint, consent boundaries are like privacy settings—they give different permissions to different sets of users. But, they differ from traditional privacy settings with respect to the categories of user-sets that can be formed, the granularity of control, and most importantly, the way the categories align with users' own preferences for control. Below, we reflect further on how consent differs from privacy, and what consent means for system development.}

\final{To begin, privacy is a fuzzy concept—scholars disagree on how exactly to define it, though definitions cluster around themes such as the right to be left alone, restriction of access, secrecy, control over personal information, personhood, and intimacy~\cite{solove2002conceptualizing}. One relevant ambiguity in the context of computing is about who provides privacy: Privacy can be conferred to a user by the user themselves, or by a privacy-protecting system.}

\final{In contrast, the essence of consent is that it is strictly user-determined. The only entity who can grant an individual's consent is the individual herself; there is no definition of ``consent'' in which a system can grant itself consent to do something to a user that the user does not herself wish to allow.} %More importantly, granting consent to a person or a system creates new moral rights for them. One example that Hurd used was—if a person consents to another person’s touch, it means they give liberty to the other person to do actions that were not morally permissible before \cite{hurd1996moral}. Analogously, when a user consents to another user’s request to tag them in a group photo on Facebook, they are granting moral permission for the user to create a digital trace of them.

\final{Thus, relying on the term “privacy” can obscure the central role of the user, as privacy encompasses multiple, often ambiguous meanings. Arguably, the reason why "user-centered privacy" is a term distinct from general privacy is because the user-centeredness of privacy is not a given. Yet, even within the usable privacy and security community, discussions of consent have been limited to the notice-and-choice paradigm \cite{cate2006failure}. Consent is not cast as a foundational principle for system design. Based on this view, systems frequently define privacy options in advance (i.e., the system is conferring some privacy), with users only able to accept or decline them. In practice, these options are often difficult to find or understand, further limiting true user consent. For instance, many of Facebook's privacy settings are often hidden \cite{habib2022identifying}. } 

\final{As the feminist affirmative consent framework recommends \cite{im2021yes},\footnote{Of course, how people express consent can be different, and there is an abundant literature on understanding such a range of expressions regarding demographic factors. Feminist scholarship on consent is a major example of this (e.g., \cite{beres2007spontaneous}). However, these differences in expression do not diminish what consent fundamentally does.} consent should not merely be a mechanism for relaxing privacy constraints, but a foundational principle of design: individuals affected by an action should have the agency to define exactly what their consent boundary is, in the way that they conceive as relevant to a particular context. }
% We observed participants making similar distinctions between agency and privacy on Moa—some Phase 3 participants explicitly stated they used consent boundaries for reasons other than privacy; what mattered most to them was expressing their preferences (see Section \ref{consent-boundary-usage-reasons}). Thus, a consent approach starts and ends with the user for all things permissions-related.  

% It is not that the field of privacy has entirely overlooked users' agency—most notably, the usable privacy community has long advocated for users' privacy rights. However, much of the discourse in usable privacy tends to focus on \emph{usability}, which partially addresses aspects of affirmative consent such as the qualities of ``informed'' and ``unburdensome'' to varying degrees (e.g., \cite{im2023less,schaub2017designing,yao2017folk}). But the consent approach goes further than the demands of usability to imply that social platforms should be designed to let people easily configure systems as they want---and this extends beyond just privacy controls. 

\subsection{Towards Collective Action} %Towards In-Person Interactions from Online Sensitive Information Sharing}

Originally, Moa was conceived to help students \emph{resolve} their advising challenges, but we recognized the problem space was very large, and narrowed our scope to ally discovery. But as reported, some users were considering next steps beyond ally discovery. A few Phase 3 participants mentioned wanting to meet another user in-person or wished it was easier to have a 1:1 interaction on Moa. 
% The need for systems like Moa exists because of power disparities: In any context, those with less power have difficulty addressing problems caused by those with more power. Limited disclosure and discussion allows the less powerful to find allies with whom to share information and to plan remedial action. Under those conditions, how could an online system facilitate next steps? 

How could a system like Moa be extended in future research?
One line of work might involve developing mechanisms that enable non-anonymous connections for collective action. While Moa has allowed PhD students to have 
conversations around advising relationships, nobody revealed their identity to other users. To make consentful non-anonymous connections easier, future work could explore embedding an information escrow into Moa so that users can gradually build trust in each other, to the point of revealing their identities to one another. Information escrows are systems that connect two parties or enable information exchange, only if both provide information (to the system) that both agree would be sufficient for trust~\cite{ayres2012information}.\footnote{Bad actors could try to submit false information to information escrows. We believe the most feasible approach to address such issues is to have human moderators. While there could be scalability issues, systems like Callisto \cite{rajan2018callisto}, show that it is feasible for escrow systems to be implemented with moderators on a scale of spanning multiple campuses. \url{https://www.projectcallisto.org/callistovault}}

However, one drawback of escrows is that they are private until the match happens---users cannot observe others, which is important for developing enough trust to submit information to escrows in the first place. We suspect it is possible to provide a middle ground by embedding information escrows within a platform like Moa, so that users can give and receive signals for reciprocal self-disclosures~
\cite{sprecher2013taking}. For instance, users can observe each other’s comments on Moa, and then decide to submit information after building a sufficient level of trust.

% First, even as some of our participants seemed ready for in-person meetings, they still expressed hesitation. After all, not everyone is who they seem online. One way to offer a final verification step is with an \emph{information escrow}~\cite{ayres2012information}, where two parties are connected to one another, or where information is exchanged, only if they both provide information (to the system) that both parties agree would be sufficient for trust. An information escrow could be enabled simply for users to express the willingness to meet: users indicate privately to the system whether they would be open to meeting with another poster/commenter in a thread, and only if there is mutual consent, does the system explicitly connect them. 

Another interesting direction is supporting users working together to arrive at a plan for action---especially in order to communicate with those in power. For a consent-based intervention to provide agency to those with less power, it helps to communicate with, or have the agreement of, parties in power---in Moa's context, faculty. We noticed some participants were thinking about this. Unprompted, some participants from both Phase 1 and 3 mentioned that Moa could be extended to convey PhD students' experiences to faculty. This requires a group of users to collectively plan and act \cite{salehi2015we,vlachokyriakos2017hci}. In particular, there are interesting questions around whether revelations about the identities of group members seem necessary, and if so, how that can be accomplished.

\subsection{{Audience Configuration and Identity Management in Sensitive Contexts}}
Some users showed consent boundaries to signal that the audience shares a common identity trait, in order for other users to chime in more comfortably. This was the most pronounced for a post that was about gender dynamics, where the poster used a women-only boundary. Most of the commenters also decided to use the same boundary for their comments and chose to display them. 

{Relatedly, a few of Phase 3 interview participants also noted that the commenting feature could be designed so that it is easier to display parts of one's identity when responding, which resembles meronymity~\cite{soliman2024mitigating}. Moa could potentially be extended to provide users with such features, so that once an audience restriction has happened, users within the thread could more freely express parts of their identity if they wanted to, separately from using consent boundaries.}

Some users who did not show their consent boundaries thought they did not add much information, since the post will be delivered to the right audience anyway. These Phase 3 participants were less concerned with expressing identity and more focused on audience control, possibly because Moa’s context centers on sensitive power dynamics, unlike prior work (e.g., \cite{soliman2024mitigating,ma2016anonymity}). For example, while meronymity~\cite{soliman2024mitigating} was evaluated where different levels of seniority exist, the focus was on academic research discourse, rather than power hierarchies.

\subsection{Scaling Moa and Consent Boundaries}
Currently, Moa is maintained by the lead author in order to minimize privacy risks. Running the platform with 47 users turned out to be feasible—the lead author was diligent in moderating and seeking user feedback, and no incidents of abuse or negative feedback occurred. However, how could Moa and the concept of consent boundaries scale?

% multiple instance
One way to scale is having multiple instances of Moa, one per organization or group, each with its own administrator, similar to Mastodon. This approach reflects an important assumption that local system admins are important for earning users’ trust, which was confirmed by our Phase 3 study. And, just as research shows that Reddit’s moderators have their own network of giving each other support~\cite{bozarth2023wisdom}, we imagine that there should \final{be} a support network for Moa's administrators and moderators. This would ensure that each instance is not entirely isolated, and that when sensitive incidents happen, there would be cross-instance resources on how to resolve them. 

\final{It is possible to imagine building Moa as a larger, monolithic platform such as Facebook or Reddit, but we imagine even on such a platform, communities would have their own moderators---each trusted by their respective communities---who are empowered to provide the kind of meaningful oversight over identity that the Moa prototype included. For this study, it was easy for users to verify and understand that the moderator was also a PhD student in the same institution. At larger scale, there would need to be a process for selecting such moderators, and proving their identity to other members of the community.}

\final{In terms of consent boundaries, they are likely to be applicable to other contexts beyond PhD advising, such as among workplace employees \cite{ramonetti2019keeping,decker2024catalyst}, survivors of sexual misbehavior by one group of people \cite{lonsway2013sexual,cortina2021putting}, or potential whistleblowers at the same organization \cite{di2022leaking}. One aspect of the Moa prototype described in this paper was that there was an implicit consent boundary formed by the community of PhD students from the same institution---no user needed to consider whether their content would be revealed to non-students. Communities could be formed much like Slack's channels around different topics, though additional mechanisms such as member-sponsored invitations to join or formal background checks would be required to ensure that community members satisfy membership requirements.}

\final{We also imagine consent boundaries could be used in general platforms such as Facebook or Reddit, likely in less sensitive contexts. Consent boundaries could help users carve out a group of people they want to interact with, without having to designate a particular space or network \cite{zhang2024form}. This would resemble the ``social circles'' features on platforms such as Close Friends,\footnote{\url{https://www.facebook.com/help/200538509990389?helpref=faq_content}} but it would not require individual enumeration of members or knowledge of other users' identities. For example, users who were exchanging comments under a post on a subreddit could use consent boundaries to easily create a space within the thread for users that meet particular criteria. At the same time, when consent boundaries are used within a large-scale commercial platform, there might be a bigger hesitation for users to use the feature. It may take a longer time for users to build trust in consent boundaries, as failures could result in unintended exposure of posts to a broad audience. Because of this, the usage rate could be much lower in the beginning.}

\section{Conclusion}
When individuals experience harm from someone in power, finding allies—people who are sympathetic or willing to navigate an issue together—is important. To understand how to develop a social platform that supports sensitive conversations for ally discovery, we conducted interviews with 19 PhD students, which revealed a strong desire for a platform that connects peers to discuss PhD advising relationships, and reinforced the importance of consent in enabling these connections. Building on these findings and the affirmative consent framework, we developed a social platform called Moa. One of Moa’s core features is the ``consent boundary,'' which allow users to fluidly demarcate who can view each post or comment based on dimensions such as social identity, lived experience, and interpersonal relationships{---while preserving mutual anonymity. The sender and recipient do not learn of each other's identity, even as the post reaches the intended audience}. We conducted a three-week field study with 47 PhD students that showed Moa successfully facilitated ally discovery. Consent boundaries lowered barriers for some users, enabling them to ask sensitive questions and provide authentic advice. In all, this work presents insights on how to develop a social platform for ally discovery and instantiates a consent-centered approach to system design.

\section{Acknowledgements}
We sincerely thank all participants. A big thank you to Vikram Mohanty for being a sounding board throughout the system development. Chaerin Im deserves a very special shout-out for encouraging Im to build a new social media platform in 2021. We also thank Jiyoon Kim, Shwetha Rajaram, and Yulin Yu for their immensely kind support and great feedback. Yunseok Jang gave much needed suggestions for the tutorial. Im also thanks Amy Ko, Mark Ackerman, and Emily Mower Provost for their very helpful comments. Im was supported by a Meta Research PhD Fellowship.

\bibliographystyle{ACM-Reference-Format}
\bibliography{references}

@article{bozarth2023wisdom,
author = {Bozarth, Lia and Im, Jane and Quarles, Christopher and Budak, Ceren},
title = {Wisdom of Two Crowds: Misinformation Moderation on Reddit and How to Improve this Process---A Case Study of COVID-19},
year = {2023},
issue_date = {April 2023},
publisher = {Association for Computing Machinery},
address = {New York, NY, USA},
volume = {7},
number = {CSCW1},
url = {https://doi.org/10.1145/3579631},
doi = {10.1145/3579631},
journal = {Proc. ACM Hum.-Comput. Interact.},
month = apr,
articleno = {155},
numpages = {33}
}

@article{cohen2022abuse,
  title={Abuse and exploitation of doctoral students: A conceptual model for traversing a long and winding road to academia},
  author={Cohen, Aaron and Baruch, Yehuda},
  journal={Journal of Business Ethics},
  volume={180},
  number={2},
  pages={505--522},
  year={2022},
  publisher={Springer}
}

@article{golde2005role,
  title={The role of the department and discipline in doctoral student attrition: Lessons from four departments},
  author={Golde, Chris M},
  journal={The Journal of Higher Education},
  volume={76},
  number={6},
  pages={669--700},
  year={2005},
  publisher={Taylor \& Francis}
}

@article{feldblum2016select,
  title={Select task force on the study of harassment in the workplace},
  author={Feldblum, Chai R and Lipnic, Victoria A},
  journal={Washington: US Equal Employment Opportunity Commission},
  year={2016},
    url={https://www.eeoc.gov/select-task-force-study-harassment-workplace#_ftn62}
}

@article{branch2013workplace,
  title={Workplace bullying, mobbing and general harassment: A review},
  author={Branch, Sara and Ramsay, Sheryl and Barker, Michelle},
  journal={International Journal of Management Reviews},
  volume={15},
  number={3},
  pages={280--299},
  year={2013},
  publisher={Wiley Online Library}
}

@inproceedings{abdulgalimov2020designing,
  title={Designing for employee voice},
  author={Abdulgalimov, Dinislam and Kirkham, Reuben and Nicholson, James and Vlachokyriakos, Vasilis and Briggs, Pam and Olivier, Patrick},
  booktitle={Proceedings of the 2020 CHI Conference on Human Factors in Computing Systems},
  pages={1--13},
  year={2020}
}

@article{solove2002conceptualizing,
  title={Conceptualizing privacy},
  author={Solove, Daniel J},
  journal={Calif. L. Rev.},
  volume={90},
  pages={1087},
  year={2002},
  publisher={HeinOnline}
}

@inproceedings{wang2016modeling,
  title={Modeling self-disclosure in social networking sites},
  author={Wang, Yi-Chia and Burke, Moira and Kraut, Robert},
  booktitle={Proceedings of the 19th ACM conference on computer-supported cooperative work \& social computing},
  pages={74--85},
  year={2016}
}

@inproceedings{wisniewski2012fighting,
  title={Fighting for my space: Coping mechanisms for SNS boundary regulation},
  author={Wisniewski, Pamela and Lipford, Heather and Wilson, David},
  booktitle={Proceedings of the SIGCHI Conference on Human Factors in Computing Systems},
  pages={609--618},
  year={2012}
}

@article{vitak2012impact,
  title={The impact of context collapse and privacy on social network site disclosures},
  author={Vitak, Jessica},
  journal={Journal of broadcasting \& electronic media},
  volume={56},
  number={4},
  pages={451--470},
  year={2012},
  publisher={Taylor \& Francis}
}

@article{collins1994self,
  title={Self-disclosure and liking: a meta-analytic review.},
  author={Collins, Nancy L and Miller, Lynn Carol},
  journal={Psychological bulletin},
  volume={116},
  number={3},
  pages={457},
  year={1994},
  publisher={American Psychological Association}
}

@misc{kim2017blind,
    title={Blind: The anonymous app that serves as Silicon Valley's gossip outlet},
    year={2017},
    author={Young-won, Kim},
    publisher={The Korean Herald},
    url={https://www.theinvestor.co.kr/view.php?ud=20170330000905}
}

@article{sprecher2013taking,
  title={Taking turns: Reciprocal self-disclosure promotes liking in initial interactions},
  author={Sprecher, Susan and Treger, Stanislav and Wondra, Joshua D and Hilaire, Nicole and Wallpe, Kevin},
  journal={Journal of Experimental Social Psychology},
  volume={49},
  number={5},
  pages={860--866},
  year={2013},
  publisher={Elsevier}
}

@article{mayer1995integrative,
  title={An integrative model of organizational trust},
  author={Mayer, Roger C and Davis, James H and Schoorman, F David},
  journal={Academy of management review},
  volume={20},
  number={3},
  pages={709--734},
  year={1995},
  publisher={Academy of Management Briarcliff Manor, NY 10510}
}

@misc{schoorman2007integrative,
  title={An integrative model of organizational trust: Past, present, and future},
  author={Schoorman, F David and Mayer, Roger C and Davis, James H},
  journal={Academy of Management review},
  volume={32},
  number={2},
  pages={344--354},
  year={2007},
  publisher={Academy of Management Briarcliff Manor, NY 10510}
}

@misc{reddit2012how,
    author={Greenberg, Andy},
    publisher={Forbes},
    year={2012},
    title={How Reddit's Alexis Ohanian Became the Mayor of the Internet},
    url={https://www.forbes.com/forbes/2012/0625/technology-politics-alexis-ohanian-reddit-sopa-mayor-of-internet.html}
}

@article{ammari2019self,
  title={Self-declared throwaway accounts on Reddit: How platform affordances and shared norms enable parenting disclosure and support},
  author={Ammari, Tawfiq and Schoenebeck, Sarita and Romero, Daniel},
  journal={Proceedings of the ACM on Human-Computer Interaction},
  volume={3},
  number={CSCW},
  pages={1--30},
  year={2019},
  publisher={ACM New York, NY, USA}
}

@inproceedings{yao2019defending,
  title={Defending my castle: A co-design study of privacy mechanisms for smart homes},
  author={Yao, Yaxing and Basdeo, Justin Reed and Kaushik, Smirity and Wang, Yang},
  booktitle={Proceedings of the 2019 chi conference on human factors in computing systems},
  pages={1--12},
  year={2019}
}

@inproceedings{powers2002privacy,
  title={Privacy promises, access control, and privacy management. Enforcing privacy throughout an enterprise by extending access control},
  author={Powers, Calvin S and Ashley, Paul and Schunter, Matthias},
  booktitle={Proceedings. Third International Symposium on Electronic Commerce,},
  pages={13--21},
  year={2002},
  organization={IEEE}
}

@misc{whittaker2018blind,
    title={At Blind, a security lapse revealed private complaints from Silicon Valley employees},
    year={2018},
    author={Whittaker, Zack},
    publisher={TechCrunch},
    url={https://techcrunch.com/2018/12/20/blind-anonymous-app-data-exposure/#:~:text=But%20Blind%20left%20one%20of,and%20identify%20would%2Dbe%20whistleblowers}
}

@article{martin2013countering,
  title={Countering supervisor exploitation},
  author={Martin, Brian},
  journal={Journal of Scholarly Publishing},
  volume={45},
  number={1},
  pages={74--86},
  year={2013},
  publisher={University of Toronto Press Incorporated}
}

@article{breen2024breaking,
  title={Breaking points: exploring how negative doctoral advisor relationships develop over time},
  author={Breen, Stephanie M and McCain, Jesse and Roksa, Josipa},
  journal={Higher Education},
  pages={1--20},
  year={2024},
  publisher={Springer}
}

@article{jhaver2019does,
  title={Does transparency in moderation really matter? User behavior after content removal explanations on reddit},
  author={Jhaver, Shagun and Bruckman, Amy and Gilbert, Eric},
  journal={Proceedings of the ACM on Human-Computer Interaction},
  volume={3},
  number={CSCW},
  pages={1--27},
  year={2019},
  publisher={ACM New York, NY, USA}
}

@inproceedings{ma2016anonymity,
  title={Anonymity, intimacy and self-disclosure in social media},
  author={Ma, Xiao and Hancock, Jeff and Naaman, Mor},
  booktitle={Proceedings of the 2016 CHI conference on human factors in computing systems},
  pages={3857--3869},
  year={2016}
}

@article{davis2014context,
  title={Context collapse: Theorizing context collusions and collisions},
  author={Davis, Jenny L and Jurgenson, Nathan},
  journal={Information, communication \& society},
  volume={17},
  number={4},
  pages={476--485},
  year={2014},
  publisher={Taylor \& Francis}
}

@article{robledo2023we,
  title={“We are Researchers, but we are also Humans”: Creating a Design Space for Managing Graduate Student Stress},
  author={Robledo Yamamoto, Fujiko and Cho, Janghee and Voida, Amy and Voida, Stephen},
  journal={ACM Transactions on Computer-Human Interaction},
  volume={30},
  number={5},
  pages={1--33},
  year={2023},
  publisher={ACM New York, NY}
}

@article{becerra2021does,
  title={Does a good advisor a day keep the doctor away? How advisor-advisee relationships are associated with psychological and physical well-being among graduate students},
  author={Becerra, Monica and Wong, Emily and Jenkins, Brooke N and Pressman, Sarah D},
  journal={International Journal of Community Well-Being},
  volume={4},
  pages={505--524},
  year={2021},
  publisher={Springer}
}

@article{peluso2011depression,
  title={Depression symptoms in Canadian psychology graduate students: Do research productivity, funding, and the academic advisory relationship play a role?},
  author={Peluso, Daniel L and Carleton, R Nicholas and Asmundson, Gordon JG},
  journal={Canadian Journal of Behavioural Science/Revue canadienne des sciences du comportement},
  volume={43},
  number={2},
  pages={119},
  year={2011},
  publisher={Educational Publishing Foundation}
}

@article{levecque2017work,
  title={Work organization and mental health problems in PhD students},
  author={Levecque, Katia and Anseel, Frederik and De Beuckelaer, Alain and Van der Heyden, Johan and Gisle, Lydia},
  journal={Research policy},
  volume={46},
  number={4},
  pages={868--879},
  year={2017},
  publisher={Elsevier}
}

@article{swazey1993ethical,
  title={Ethical problems in academic research},
  author={Swazey, Judith P and Anderson, Melissa S and Lewis, Karen Seashore and Louis, Karen Seashore},
  journal={American Scientist},
  volume={81},
  number={6},
  pages={542--553},
  year={1993},
  publisher={JSTOR}
}

@inproceedings{im2021yes,
author = {Im, Jane and Dimond, Jill and Berton, Melody and Lee, Una and Mustelier, Katherine and Ackerman, Mark S. and Gilbert, Eric},
title = {Yes: Affirmative Consent as a Theoretical Framework for Understanding and Imagining Social Platforms},
year = {2021},
isbn = {9781450380966},
publisher = {Association for Computing Machinery},
address = {New York, NY, USA},
booktitle = {Proceedings of the 2021 CHI Conference on Human Factors in Computing Systems},
articleno = {403},
numpages = {18},
location = {Yokohama, Japan},
series = {CHI '21},
url={https://doi.org/10.1145/3411764.3445778}
}

@article{zhang2024form,
  title={Form-From: A Design Space of Social Media Systems},
  author={Zhang, Amy X and Bernstein, Michael S and Karger, David R and Ackerman, Mark S},
  journal={Proceedings of the ACM on Human-Computer Interaction},
  volume={8},
  number={CSCW1},
  pages={1--47},
  year={2024},
  publisher={ACM New York, NY, USA}
}

@inproceedings{soliman2024mitigating,
  title={Mitigating Barriers to Public Social Interaction with Meronymous Communication},
  author={Soliman, Nouran and Kang, Hyeonsu B and Latzke, Matthew and Bragg, Jonathan and Chang, Joseph Chee and Zhang, Amy Xian and Karger, David R},
  booktitle={Proceedings of the CHI Conference on Human Factors in Computing Systems},
  pages={1--26},
  year={2024}
}

@article{hurd1996moral,
  title={The moral magic of consent},
  author={Hurd, Heidi M},
  journal={Legal theory},
  volume={2},
  number={2},
  pages={121--146},
  year={1996},
  publisher={Cambridge University Press}
}

@article{tilly2004social,
  title={Social boundary mechanisms},
  author={Tilly, Charles},
  journal={Philosophy of the social sciences},
  volume={34},
  number={2},
  pages={211--236},
  year={2004},
  publisher={Sage Publications}
}

@article{kim2025trust,
  title={Trust-Enabled Privacy: Social Media Designs to Support Adolescent User Boundary Regulation},
  author={Kim, JaeWon and Wolfe, Robert and Subramanian, Ramya Bhagirathi and Lee, Mei-Hsuan and Colnago, Jessica and Hiniker, Alexis},
  journal={arXiv preprint arXiv:2502.19082},
  year={2025}
}

@inproceedings{grandhi2016reply,
  title={To reply or to reply all: Understanding replying behavior in group email communication},
  author={Grandhi, Sukeshini A and Lanagan-Leitzel, Lyndsey K},
  booktitle={Proceedings of the 19th ACM Conference on Computer-Supported Cooperative Work \& Social Computing},
  pages={560--569},
  year={2016}
}

@article{skovholt2006email,
  title={Email copies in workplace interaction},
  author={Skovholt, Karianne and Svennevig, Jan},
  journal={Journal of Computer-Mediated Communication},
  volume={12},
  number={1},
  pages={42--65},
  year={2006},
  publisher={Oxford University Press Oxford, UK}
}

@article{machili2019snowball,
  title={‘The snowball of emails we deal with’: CCing in multinational companies},
  author={Machili, Ifigeneia and Angouri, Jo and Harwood, Nigel},
  journal={Business and Professional Communication Quarterly},
  volume={82},
  number={1},
  pages={5--37},
  year={2019},
  publisher={SAGE Publications Sage CA: Los Angeles, CA}
}

@article{haesevoets2021transparency,
  title={Transparency and control in email communication: The more the supervisor is put in cc the less trust is felt},
  author={Haesevoets, Tessa and De Cremer, David and De Schutter, Leander and McGuire, Jack and Yang, Yu and Jian, Xie and Van Hiel, Alain},
  journal={Journal of Business Ethics},
  volume={168},
  pages={733--753},
  year={2021},
  publisher={Springer}
}

@article{paci2018survey,
  title={Survey on access control for community-centered collaborative systems},
  author={Paci, Federica and Squicciarini, Anna and Zannone, Nicola},
  journal={ACM Computing Surveys (CSUR)},
  volume={51},
  number={1},
  pages={1--38},
  year={2018},
  publisher={ACM New York, NY, USA}
}

@article{zhang2020configuring,
  title={Configuring audiences: A case study of email communication},
  author={Zhang, Justine and Pennebaker, James and Dumais, Susan and Horvitz, Eric},
  journal={Proceedings of the ACM on Human-Computer Interaction},
  volume={4},
  number={CSCW1},
  pages={1--26},
  year={2020},
  publisher={ACM New York, NY, USA}
}

@book{kahan2014theories,
  title={Theories of coalition formation},
  author={Kahan, James P and Rapoport, Amnon},
  year={2014},
  publisher={Psychology Press}
}

@article{mcquaid2021surviving,
  title={Surviving violent, traumatic loss after severe political persecution: lessons from the evaluation of a Venezuelan asylum seeker},
  author={McQuaid, Jennifer H and Silva, Michelle Alejandra and McKenzie, Katherine C},
  journal={BMJ Case Reports CP},
  volume={14},
  number={3},
  pages={e239025},
  year={2021},
  publisher={BMJ Specialist Journals}
}

@article{fullerrankism,
  title={Rankism: A Social Disorder},
  author={Fuller, Robert W},
  journal={Breaking Ranks}
}

@article{ayres2012information,
  title={Information escrows},
  author={Ayres, Ian and Unkovic, Cait},
  journal={Mich. L. Rev.},
  volume={111},
  pages={145},
  year={2012},
  publisher={HeinOnline}
}

@article{zapf2020empirical,
  title={Empirical findings on prevalence and risk groups of bullying in the workplace},
  author={Zapf, Dieter and Escart{\'\i}n, Jordi and Scheppa-Lahyani, Miriam and Einarsen, St{\aa}le Valvatne and Hoel, Helge and Vartia, Maarit},
  journal={Bullying and harassment in the workplace: Theory, research and practice},
  volume={3},
  pages={105--162},
  year={2020},
  publisher={CRC Press Boca Raton, FL}
}

@inproceedings{vlachokyriakos2017hci,
  title={HCI, solidarity movements and the solidarity economy},
  author={Vlachokyriakos, Vasillis and Crivellaro, Clara and Wright, Pete and Karamagioli, Evika and Staiou, Eleni-Revekka and Gouscos, Dimitris and Thorpe, Rowan and Kr{\"u}ger, Antonio and Sch{\"o}ning, Johannes and Jones, Matt and others},
  booktitle={Proceedings of the 2017 CHI conference on human factors in computing systems},
  pages={3126--3137},
  year={2017}
}

@article{lothian2021archive,
  title={An archive of whose own? White feminism and racial justice in fan fiction's digital infrastructure},
  author={Lothian, Alexis and Stanfill, Mel},
  journal={Transformative Works and Cultures},
  volume={36},
  pages={2021},
  year={2021}
}

@article{di2022leaking,
  title={Leaking black boxes: Whistleblowing and big tech invisibility},
  author={Di Salvo, Philip},
  journal={First Monday},
  year={2022}
}

@article{cortina2021putting,
  title={Putting people down and pushing them out: Sexual harassment in the workplace},
  author={Cortina, Lilia M and Areguin, Maira A},
  journal={Annual Review of Organizational Psychology and Organizational Behavior},
  volume={8},
  number={1},
  pages={285--309},
  year={2021},
  publisher={Annual Reviews}
}

@article{beres2007spontaneous,
  title={`Spontaneous' sexual consent: An analysis of sexual consent literature},
  author={Beres, Melanie A},
  journal={Feminism \& Psychology},
  volume={17},
  number={1},
  pages={93--108},
  year={2007},
  publisher={Sage Publications Sage CA: Thousand Oaks, CA},
  doi={10.1177/0959353507072914},
  url={https://doi.org/10.1177/0959353507072914}
}

@article{muehlenhard2016complexities,
  title={The complexities of sexual consent among college students: A conceptual and empirical review},
  author={Muehlenhard, Charlene L and Humphreys, Terry P and Jozkowski, Kristen N and Peterson, Zo{\"e} D},
  journal={The Journal of Sex Research},
  volume={53},
  number={4-5},
  pages={457--487},
  year={2016},
  publisher={Taylor \& Francis},
  url={https://doi.org/10.1080/00224499.2016.1146651 },
  doi={10.1080/00224499.2016.1146651}
}

@inproceedings{bernstein2013quantifying,
  title={Quantifying the invisible audience in social networks},
  author={Bernstein, Michael S and Bakshy, Eytan and Burke, Moira and Karrer, Brian},
  booktitle={Proceedings of the SIGCHI conference on human factors in computing systems},
  pages={21--30},
  year={2013}
}

@article{jones2023train,
  title={How to train your algorithm: The struggle for public control over private audience commodities on Tiktok},
  author={Jones, Corinne},
  journal={Media, Culture \& Society},
  volume={45},
  number={6},
  pages={1192--1209},
  year={2023},
  publisher={SAGE Publications Sage UK: London, England}
}

@article{page2019pragmatic,
  title={Pragmatic tool vs. relational hindrance: Exploring why some social media users avoid privacy features},
  author={Page, Xinru and Ghaiumy Anaraky, Reza and Knijnenburg, Bart P and Wisniewski, Pamela J},
  journal={Proceedings of the ACM on Human-Computer Interaction},
  volume={3},
  number={CSCW},
  pages={1--23},
  year={2019},
  publisher={ACM New York, NY, USA}
}

@article{kim2024respect,
  title={ReSPect: Enabling Active and Scalable Responses to Networked Online Harassment},
  author={Kim, Haesoo and Lee, Juhoon and Jang, Jeong-Woo and Kim, Juho},
  journal={Proceedings of the ACM on Human-Computer Interaction},
  volume={8},
  number={CSCW1},
  pages={1--30},
  year={2024},
  publisher={ACM New York, NY, USA}
}

@article{baughan2024supporting,
  title={Supporting Hard Conversations in Close Relationships Through Design},
  author={Baughan, Amanda and Tian, Larry and Shekar, Pranav and Zhang, Amy and Hiniker, Alexis},
  journal={Proceedings of the ACM on Human-Computer Interaction},
  volume={8},
  number={CSCW2},
  pages={1--22},
  year={2024},
  publisher={ACM New York, NY, USA}
}

@article{marwick2011tweet,
  title={I tweet honestly, I tweet passionately: Twitter users, context collapse, and the imagined audience},
  author={Marwick, Alice E and Boyd, Danah},
  journal={New media \& society},
  volume={13},
  number={1},
  pages={114--133},
  year={2011},
  publisher={Sage Publications Sage UK: London, England},
  url={https://doi.org/10.1177/1461444810365313 },
  doi={10.1177/1461444810365313}
}

@inproceedings{im2023less,
  title={Less is Not More: Improving Findability and Actionability of Privacy Controls for Online Behavioral Advertising},
  author={Im, Jane and Wang, Ruiyi and Lyu, Weikun and Cook, Nick and Habib, Hana and Cranor, Lorrie Faith and Banovic, Nikola and Schaub, Florian},
  booktitle={Proceedings of the 2023 CHI Conference on Human Factors in Computing Systems},
  pages={1--33},
  year={2023}
}

@inproceedings{seering2017shaping,
  title={Shaping pro and anti-social behavior on twitch through moderation and example-setting},
  author={Seering, Joseph and Kraut, Robert and Dabbish, Laura},
  booktitle={Proceedings of the 2017 ACM conference on computer supported cooperative work and social computing},
  pages={111--125},
  year={2017}
}

@article{barnes2009nature,
  title={The nature of exemplary doctoral advisors' expectations and the ways they may influence doctoral persistence},
  author={Barnes, Benita J},
  journal={Journal of College Student Retention: Research, Theory \& Practice},
  volume={11},
  number={3},
  pages={323--343},
  year={2009},
  publisher={SAGE Publications Sage CA: Los Angeles, CA}
}

@book{lee2017consent,
  title={Building Consentful Tech},
  author={Lee, Una and Toliver, Dann},
  year={2017},
  url={http://www.consentfultech.io/wp-content/uploads/2019/10/Building-Consentful-Tech.pdf}
}

@techreport{jacob2018aveth,
  title={AVETH survey on supervision of doctoral students},
  author={Jacob, Romain and Kuzmanovska, Irena and Ripin, Nina},
  year={2018},
  institution={ETH Zurich}
}

@article{woolston2017graduate,
  title={Graduate survey: A love--hurt relationship},
  author={Woolston, Chris},
  journal={Nature},
  volume={550},
  number={7677},
  pages={549--552},
  year={2017},
  publisher={Nature Publishing Group UK London}
}

@article{friedensen2023power,
  title={Power-conscious ecosystems: Understanding how power dynamics in US doctoral advising shape students’ experiences},
  author={Friedensen, Rachel E and Bettencourt, Genia M and Bartlett, Megan L},
  journal={Higher Education},
  pages={1--16},
  year={2023},
  publisher={Springer}
}

@article{halley2016move,
  title={The move to affirmative consent},
  author={Halley, Janet},
  journal={Signs: Journal of Women in Culture and Society},
  volume={42},
  number={1},
  pages={257--279},
  year={2016},
  publisher={University of Chicago Press Chicago, IL},
  doi={10.1086/686904},
  url={https://doi.org/10.1086/686904 }
}

@article{decker2024catalyst,
  title={A catalyst for activation against racism: case study on effectiveness of workplace equity, inclusion and diversity conversations},
  author={Decker-Tonnesen, Patrick and Kamunga, Kabuika and Garcia, Erick and Ibarra, Monica and Martin, Isabelle and Saliba, Kara and Beards, Caleta and Jordan, Barbara and Bhagra, Anjali},
  journal={Journal of Workplace Learning},
  year={2024},
  publisher={Emerald Publishing Limited}
}

@article{ramonetti2019keeping,
  title={Keeping the equity, inclusion, and diversity conversations going},
  author={Ramonetti, Mona and Pilato, Victoria},
  journal={Urban Library Journal},
  volume={25},
  number={1},
  pages={1},
  year={2019}
}

@article{bryden1999redefining,
  title={Redefining rape},
  author={Bryden, David P},
  journal={Buff. Crim. L. Rev.},
  volume={3},
  pages={317},
  year={1999},
  publisher={HeinOnline}
}

@article{wertheimer1999consent,
  title={What Is Consent and Is It Important?},
  author={Wertheimer, Alan},
  journal={Buff. Crim. L. Rev.},
  volume={3},
  pages={557},
  year={1999},
  publisher={HeinOnline}
}

@article{donath2007signals,
  title={Signals in social supernets},
  author={Donath, Judith},
  journal={Journal of Computer-Mediated Communication},
  volume={13},
  number={1},
  pages={231--251},
  year={2007},
  publisher={Oxford University Press Oxford, UK},
  doi={10.1111/j.1083-6101.2007.00394.x},
  url={https://doi.org/10.1111/j.1083-6101.2007.00394.x }
}

@inproceedings{mondal2016forgetting,
  title={Forgetting in social media: Understanding and controlling longitudinal exposure of socially shared data},
  author={Mondal, Mainack and Messias, Johnnatan and Ghosh, Saptarshi and Gummadi, Krishna P and Kate, Aniket},
  booktitle={Twelfth Symposium on Usable Privacy and Security ($\{$SOUPS$\}$ 2016)},
  pages={287--299},
  year={2016}
}

@article{habib2022identifying,
author = {Habib, Hana and Pearman, Sarah and Young, Ellie and Saxena, Ishika and Zhang, Robert and Cranor, Lorrie FaIth},
title = {Identifying User Needs for Advertising Controls on Facebook},
year = {2022},
issue_date = {April 2022},
publisher = {Association for Computing Machinery},
address = {New York, NY, USA},
volume = {6},
number = {CSCW1},
url = {https://doi.org/10.1145/3512906},
doi = {10.1145/3512906},
journal = {Proc. ACM Hum.-Comput. Interact.},
month = {apr},
articleno = {59},
numpages = {42}
}

@book{friedman2019yes,
  title={Yes means yes!: Visions of female sexual power and a world without rape},
  author={Friedman, Jaclyn and Valenti, Jessica},
  year={2019},
  publisher={Seal Press}
}

@article{obar2020biggest,
  title={The biggest lie on the internet: Ignoring the privacy policies and terms of service policies of social networking services},
  author={Obar, Jonathan A and Oeldorf-Hirsch, Anne},
  journal={Information, Communication \& Society},
  volume={23},
  number={1},
  pages={128--147},
  year={2020},
  publisher={Taylor \& Francis}
}

@article{solove2012introduction,
  title={Introduction: Privacy self-management and the consent dilemma},
  author={Solove, Daniel J},
  journal={Harv. L. Rev.},
  volume={126},
  pages={1880},
  year={2012},
  publisher={HeinOnline}
}

@article{madden2017privacy,
  title={Privacy, poverty, and big data: A matrix of vulnerabilities for poor Americans},
  author={Madden, Mary and Gilman, Michele and Levy, Karen and Marwick, Alice},
  journal={Wash. UL Rev.},
  volume={95},
  pages={53},
  year={2017},
  publisher={HeinOnline}
}

@article{reidenberg2015disagreeable,
  title={Disagreeable privacy policies: Mismatches between meaning and users' understanding},
  author={Reidenberg, Joel R and Breaux, Travis and Cranor, Lorrie Faith and French, Brian and Grannis, Amanda and Graves, James T and Liu, Fei and McDonald, Aleecia and Norton, Thomas B and Ramanath, Rohan},
  journal={Berkeley Tech. LJ},
  volume={30},
  pages={39},
  year={2015},
  publisher={HeinOnline}
}

@article{cate2006failure,
  title={The failure of fair information practice principles},
  author={Cate, Fred H},
  journal={Consumer protection in the age of the information economy},
  year={2006}
}

@inproceedings{habib2021toggles,
  title={Toggles, dollar signs, and triangles: How to (in) effectively convey privacy choices with icons and link texts},
  author={Habib, Hana and Zou, Yixin and Yao, Yaxing and Acquisti, Alessandro and Cranor, Lorrie and Reidenberg, Joel and Sadeh, Norman and Schaub, Florian},
  booktitle={Proceedings of the 2021 CHI Conference on Human Factors in Computing Systems},
  pages={1--25},
  year={2021}
}

@article{nissenbaum2011contextual,
  title={A contextual approach to privacy online},
  author={Nissenbaum, Helen},
  journal={Daedalus},
  volume={140},
  number={4},
  pages={32--48},
  year={2011},
  publisher={MIT Press One Rogers Street, Cambridge, MA 02142-1209, USA journals-info~…}
}

@book{schaub2020usable,
  title={Usable and Useful Privacy Interfaces},
  author={Schaub, Florian and Cranor, Lorrie},
  year={2020},
  publisher={IAPP}
}

@inproceedings{feng2021design,
  title={A design space for privacy choices: Towards meaningful privacy control in the internet of things},
  author={Feng, Yuanyuan and Yao, Yaxing and Sadeh, Norman},
  booktitle={Proceedings of the 2021 CHI Conference on Human Factors in Computing Systems},
  pages={1--16},
  year={2021}
}

@article{lowens2025misalignments,
  title={Misalignments and Demographic Differences in Expected and Actual Privacy Settings on Facebook},
  author={Lowens, Byron and Scarnecchia, Sean and Im, Jane and Afnan, Tanisha and Chen, Annie and Zou, Yixin and Schaub, Florian},
  journal={Proceedings on Privacy Enhancing Technologies},
  year={2025}
}

@article{kinnersley2013interventions,
  title={Interventions to promote informed consent for patients undergoing surgical and other invasive healthcare procedures},
  author={Kinnersley, Paul and Phillips, Katie and Savage, Katherine and Kelly, Mark J and Farrell, Elinor and Morgan, Ben and Whistance, Robert and Lewis, Vicky and Mann, Mala K and Stephens, Bethan L and others},
  journal={Cochrane Database of Systematic Reviews},
  number={7},
  year={2013},
  publisher={John Wiley \& Sons, Ltd}
}

@article{lonsway2013sexual,
  title={Sexual harassment in law enforcement: Incidence, impact, and perception},
  author={Lonsway, Kimberly A and Paynich, Rebecca and Hall, Jennifer N},
  journal={Police Quarterly},
  volume={16},
  number={2},
  pages={177--210},
  year={2013},
  publisher={Sage Publications Sage CA: Los Angeles, CA}
}

@article{fox2017metoo,
  title={\# MeToo’s global moment: The anatomy of a viral campaign},
  author={Fox, Kara and Diehm, Jan},
  journal={CNN, November},
  volume={9},
  year={2017}
}

@inproceedings{rajan2018callisto,
  title={Callisto: A cryptographic approach to detecting serial perpetrators of sexual misconduct},
  author={Rajan, Anjana and Qin, Lucy and Archer, David W and Boneh, Dan and Lepoint, Tancrede and Varia, Mayank},
  booktitle={Proceedings of the 1st ACM SIGCAS Conference on Computing and Sustainable Societies},
  pages={1--4},
  year={2018}
}

@inproceedings{leavitt2015throwaway,
  title={This is a Throwaway Account" Temporary Technical Identities and Perceptions of Anonymity in a Massive Online Community},
  author={Leavitt, Alex},
  booktitle={Proceedings of the 18th ACM conference on computer supported cooperative work \& social computing},
  pages={317--327},
  year={2015}
}

@article{resnick2012starting,
  title={Starting new online communities},
  author={Resnick, Paul and Konstan, Joseph and Chen, Yan and Kraut, Robert E},
  journal={Building successful online communities: Evidence-based social design},
  volume={231},
  year={2012},
  publisher={MIT Press, Cambridge}
}

@inproceedings{salehi2015we,
  title={We are dynamo: Overcoming stalling and friction in collective action for crowd workers},
  author={Salehi, Niloufar and Irani, Lilly C and Bernstein, Michael S and Alkhatib, Ali and Ogbe, Eva and Milland, Kristy and Clickhappier},
  booktitle={Proceedings of the 33rd annual ACM conference on human factors in computing systems},
  pages={1621--1630},
  year={2015}
}

@article{lofstrom2020ethics,
  title={What are ethics in doctoral supervision, and how do they matter? Doctoral students’ perspective},
  author={L{\"o}fstr{\"o}m, Erika and Pyh{\"a}lt{\"o}, Kirsi},
  journal={Scandinavian Journal of Educational Research},
  volume={64},
  number={4},
  pages={535--550},
  year={2020},
  publisher={Taylor \& Francis}
}

\appendix

\clearpage
\section{Appendix}
\subsection{Participants' demographics for Phase 3 study}
55.3\% were international students, 38.3\% were domestic, and 6.4\% did not submit any relevant information.
48.9\% of the users were women, 31.9\% were men, 2.1\% were non-binary, and 17\% did not submit gender information. 65.9\% were Asian, 14.9\% were Caucasian, 8.51\% were Middle Eastern or North African, 4.26\% were of Hispanic, Latinx, or Spanish origin, and 2.13\% were Black or African American. 4.2\% had mixed race and 6.4\% did not submit information. 

\begin{figure}[b!]
    \centering
     \includegraphics[width=.6\linewidth]{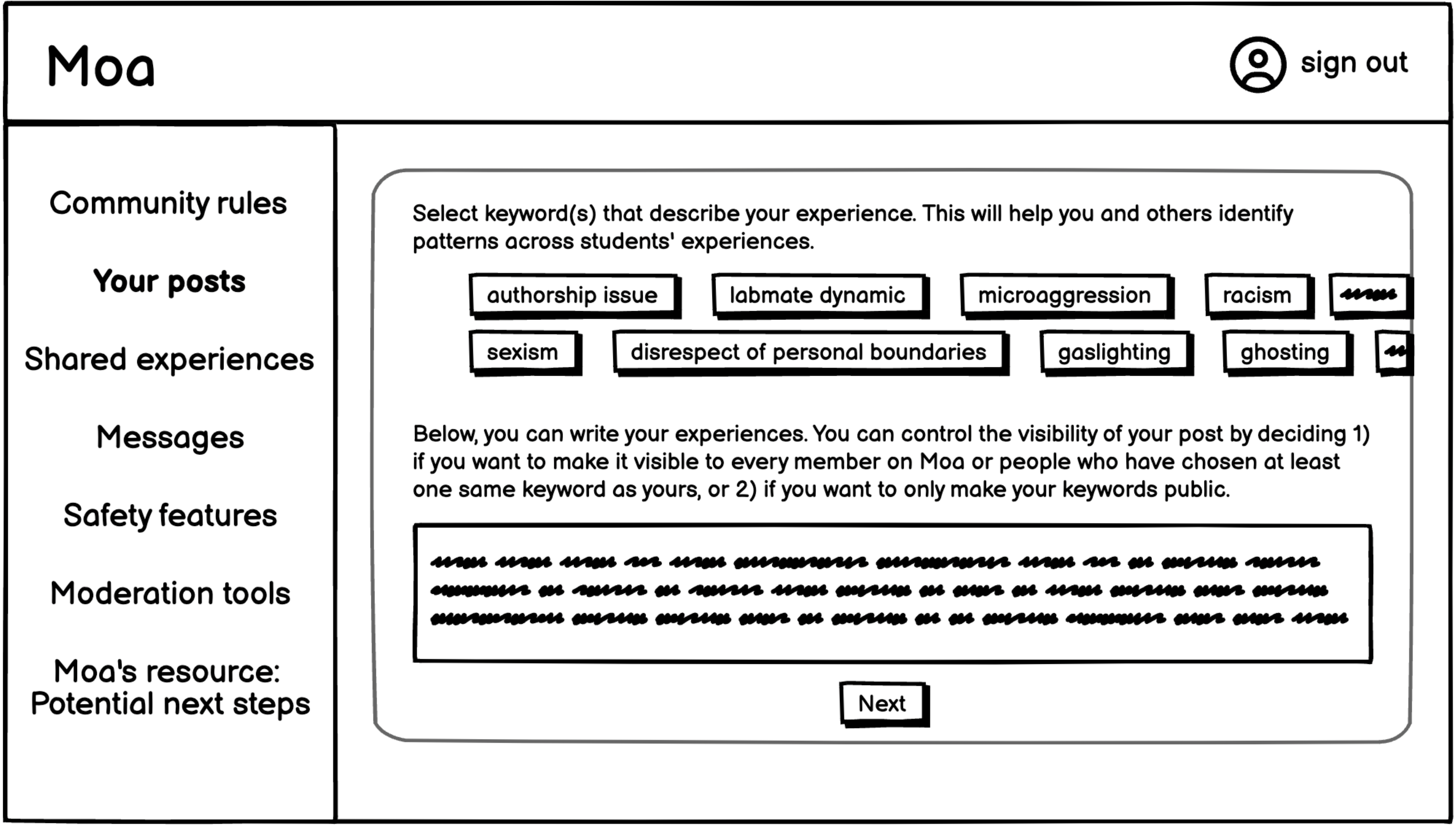}
     \includegraphics[width=.6\linewidth]{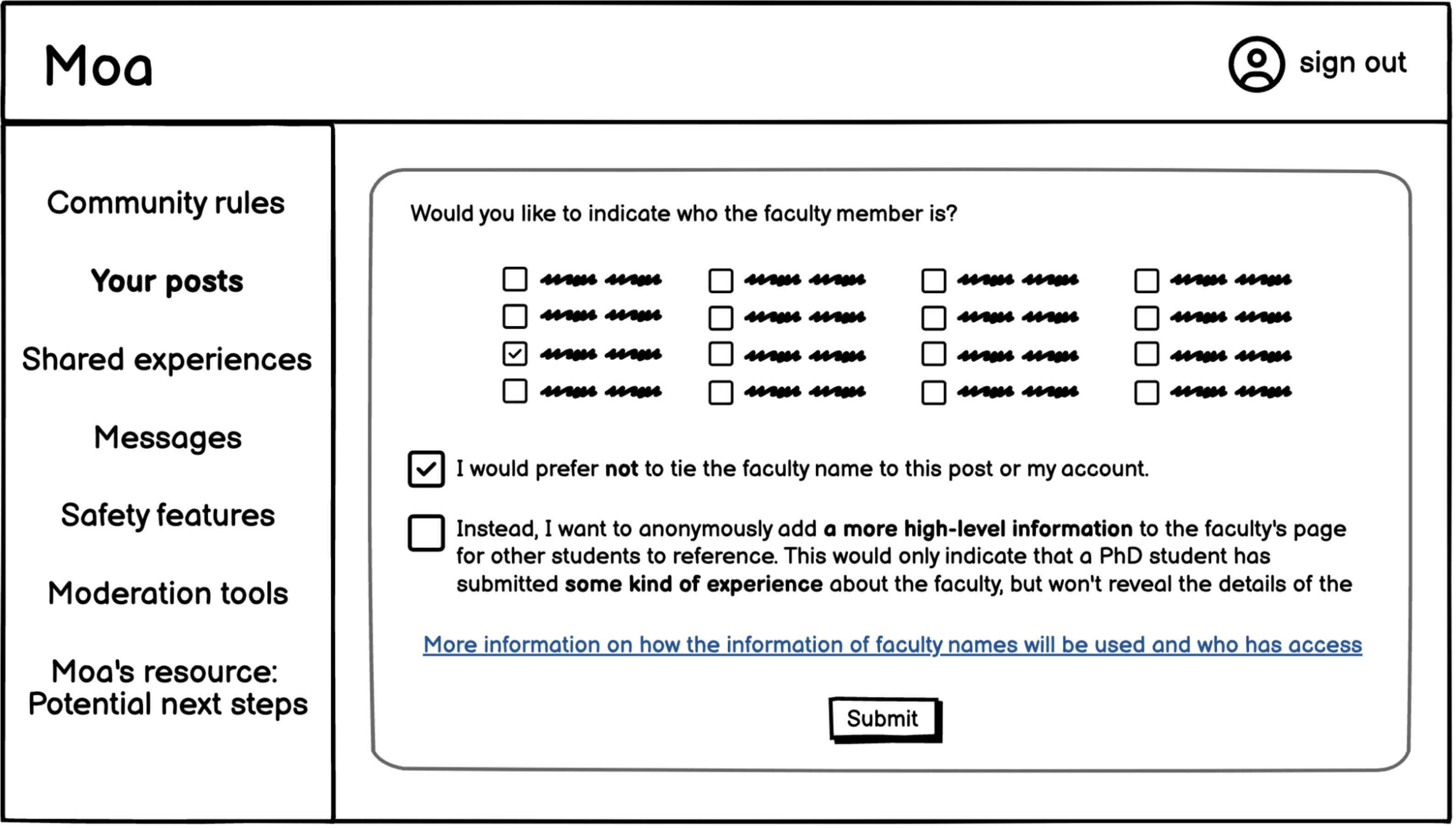}
     \includegraphics[width=.6\linewidth]{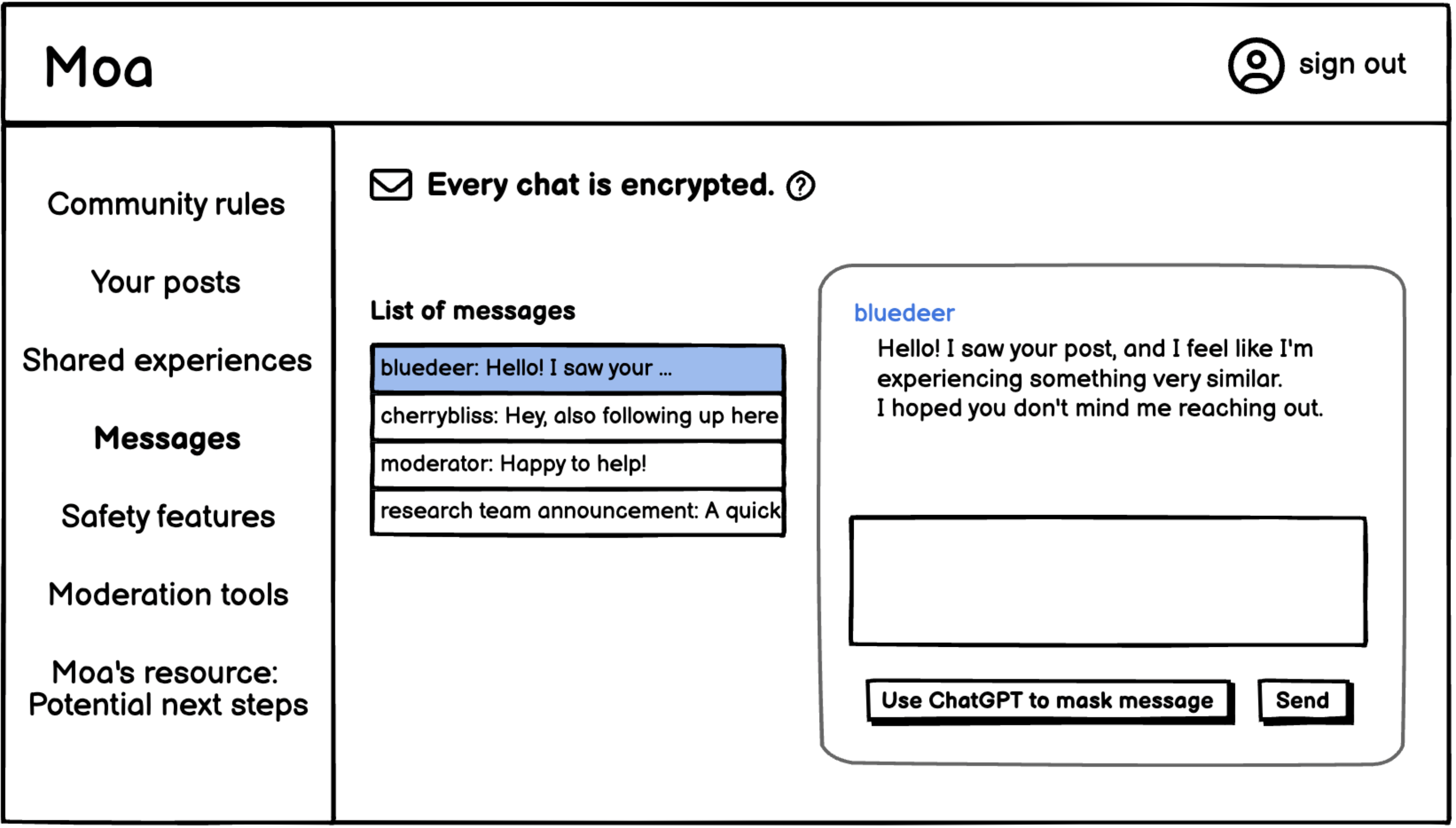}
    \vspace{10px}
\caption{Example designs that show a system for enabling PhD students to connect around advising challenges.}
\label{formative-study-designs}
\end{figure}

% \begin{figure}
%     \centering
%    \includegraphics[width=.95\linewidth]{figures/prior-pilot.png} 
%     \caption{Original consent boundary interface before the pilot studies.}
%     \label{boundary-oldversion}
% \end{figure}

\begin{figure}
    \centering
     \includegraphics[width=.95\linewidth]{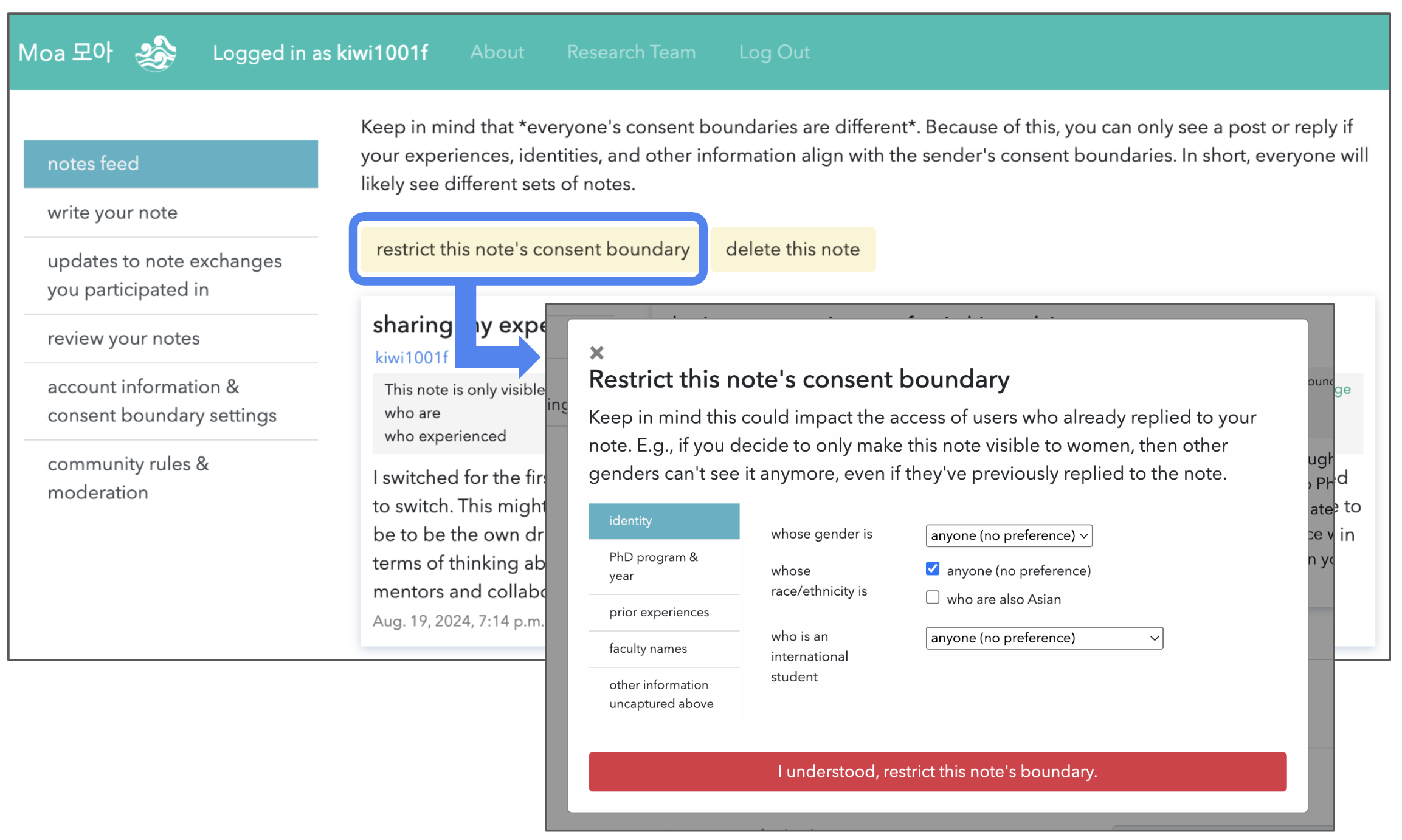}
    \caption{Users can restrict a post or comment's consent boundary.}
    \label{restrict-boundary-screenshot}
\end{figure}

\begin{figure*}[t!]
    \includegraphics[width=1\linewidth]{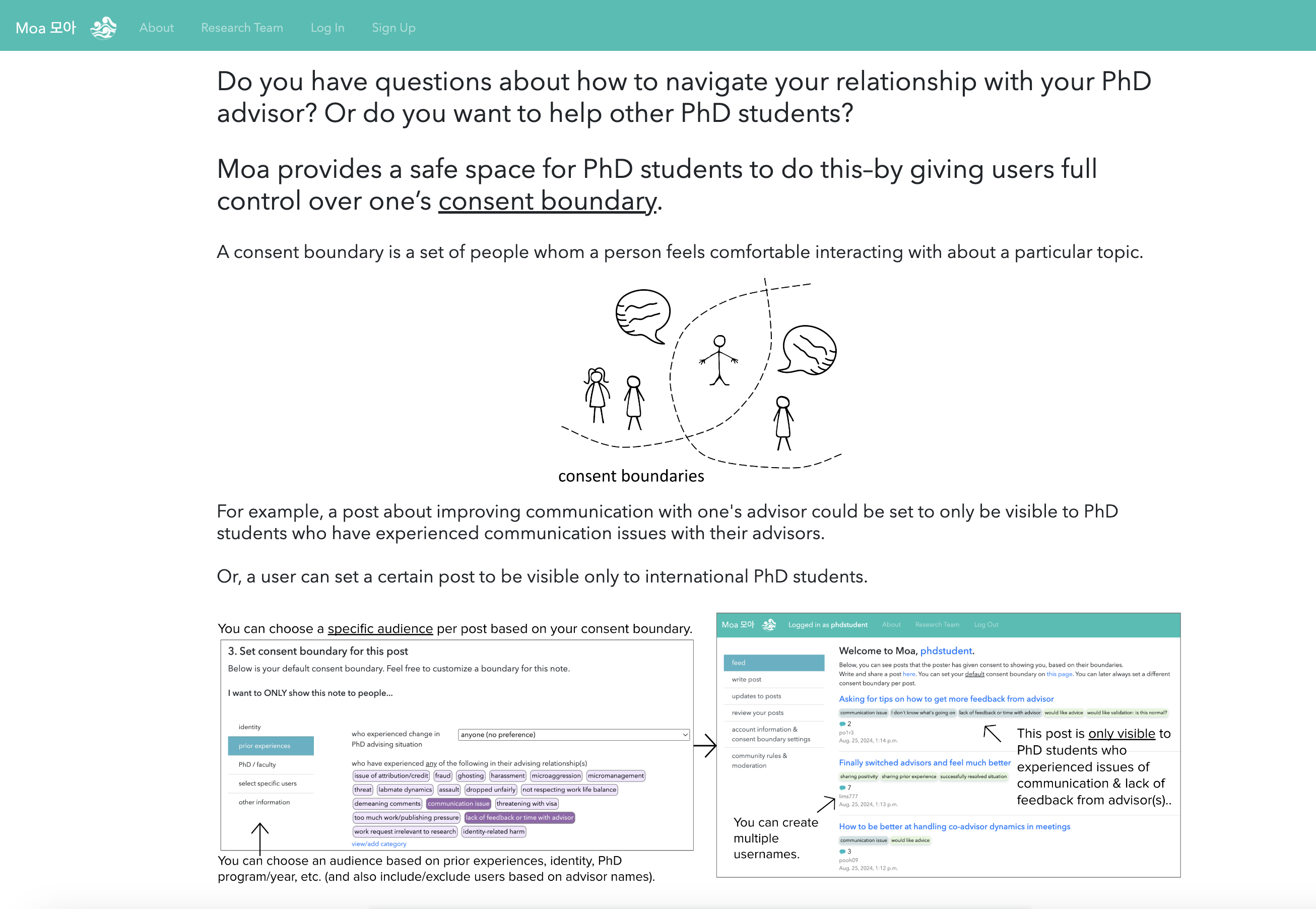}
    \caption{Moa's landing page.}
    \label{fig:moa-landingpage}
\end{figure*}

\begin{figure*}[t!]
    \includegraphics[width=0.65\linewidth]{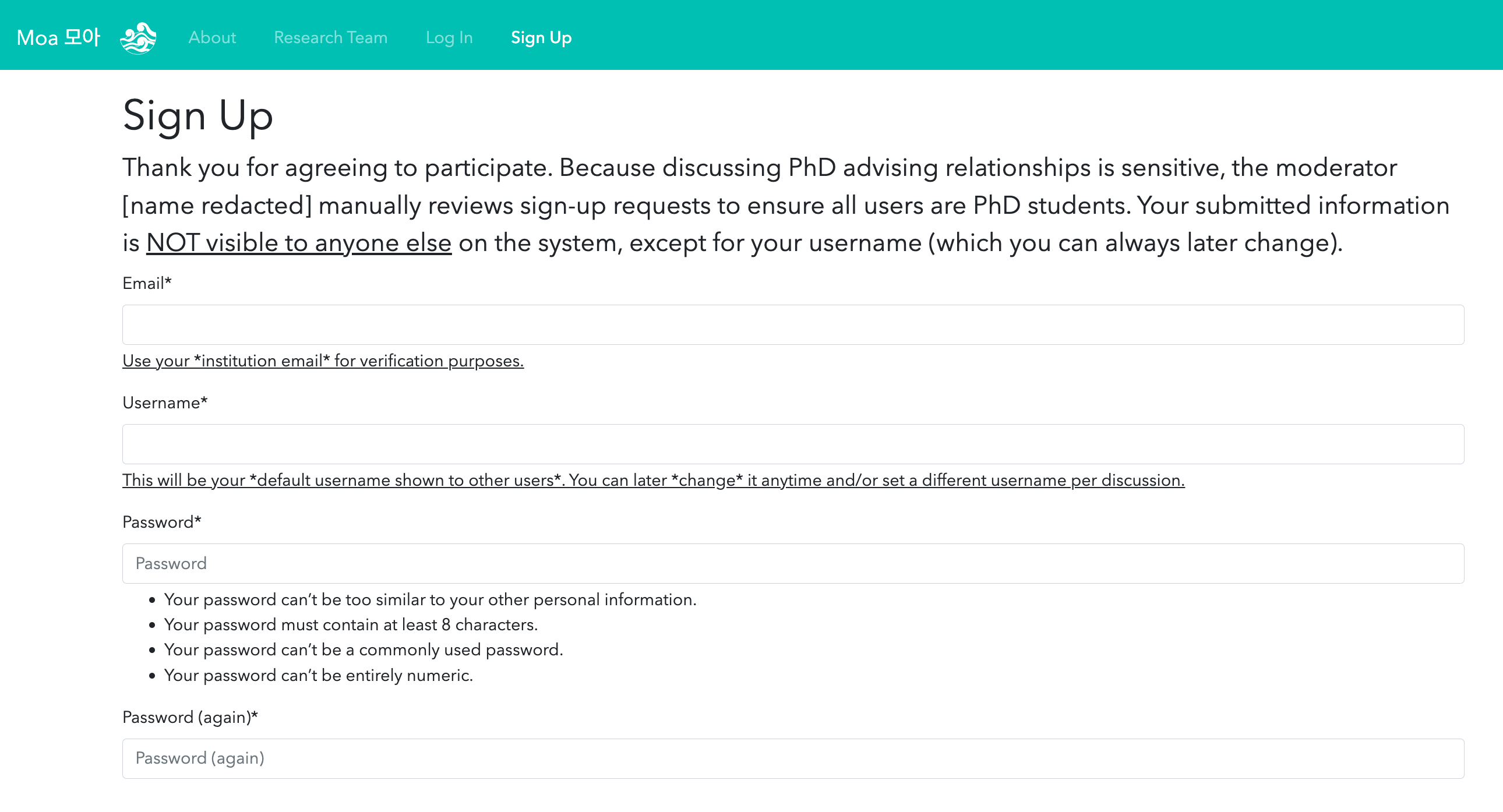}
    
     \vspace{15px}
     \includegraphics[width=0.65\linewidth]{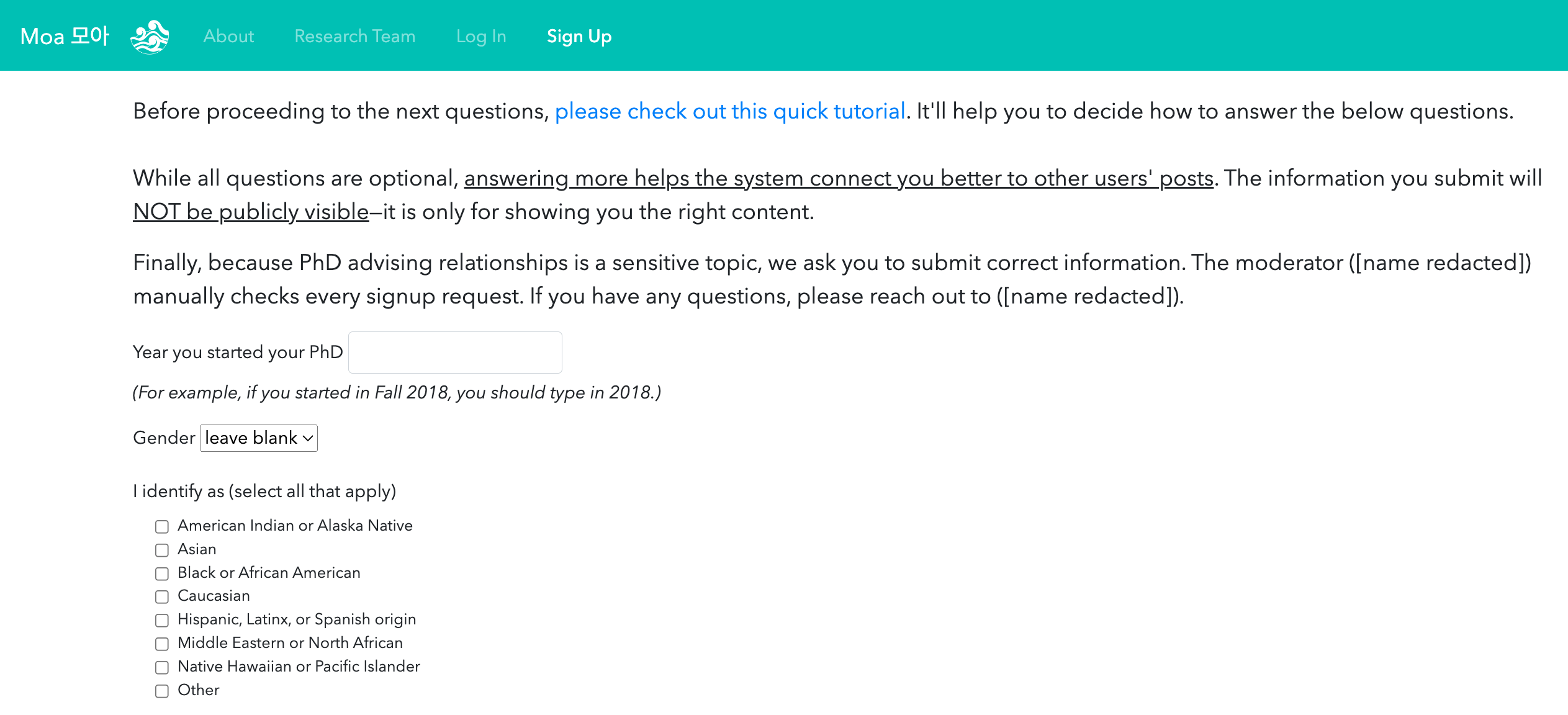}
     
    \vspace{15px}
    \includegraphics[width=0.65\linewidth]{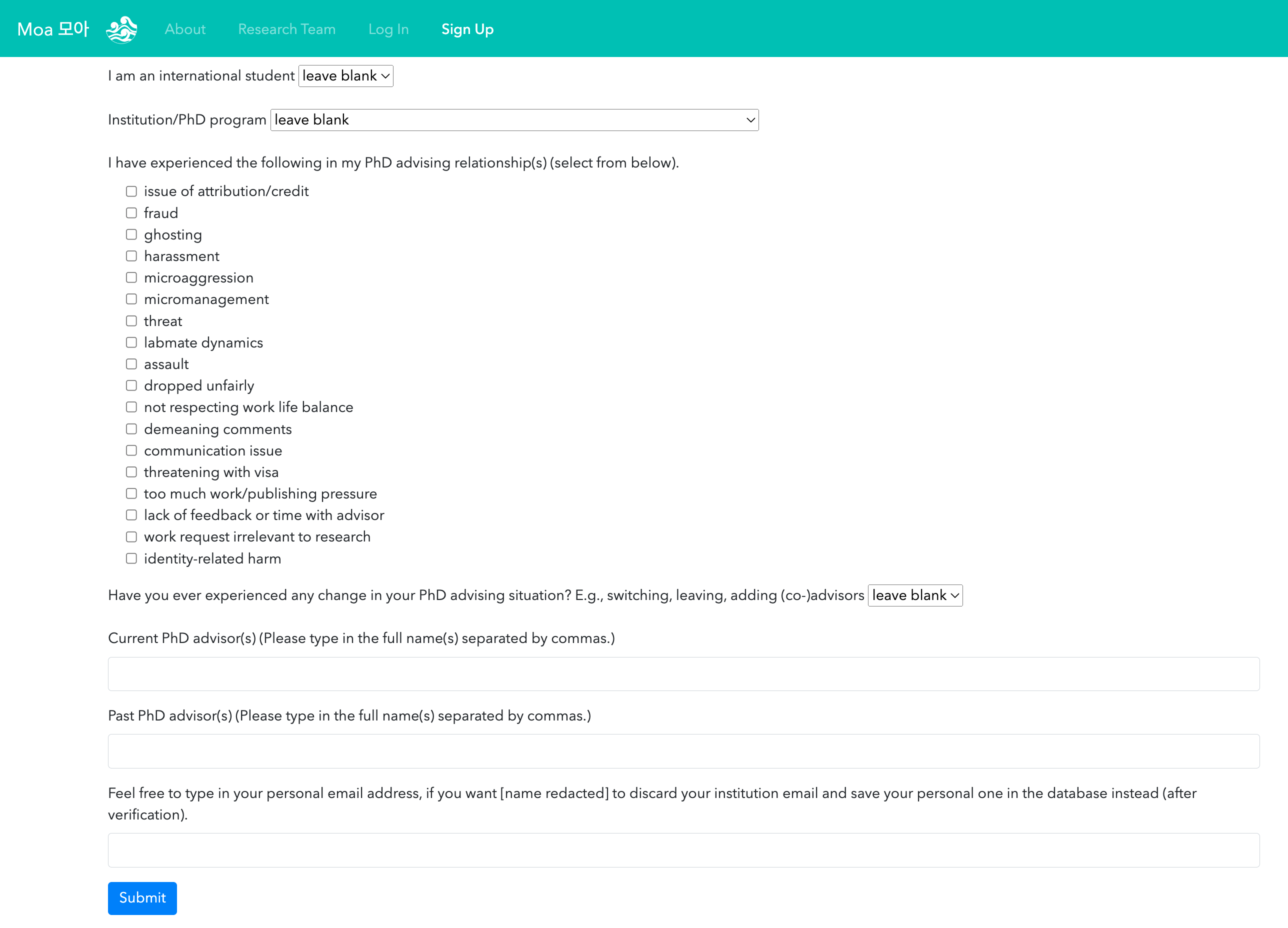}
    \caption{Moa's sign-up page. Users are asked to \textit{voluntarily} submit identity-related information, PhD program, advisor names, and challenges experienced in advising relationships.}
    \label{fig:moa-signup}
\end{figure*}

\begin{figure*}[t!]
    \includegraphics[width=1\linewidth]{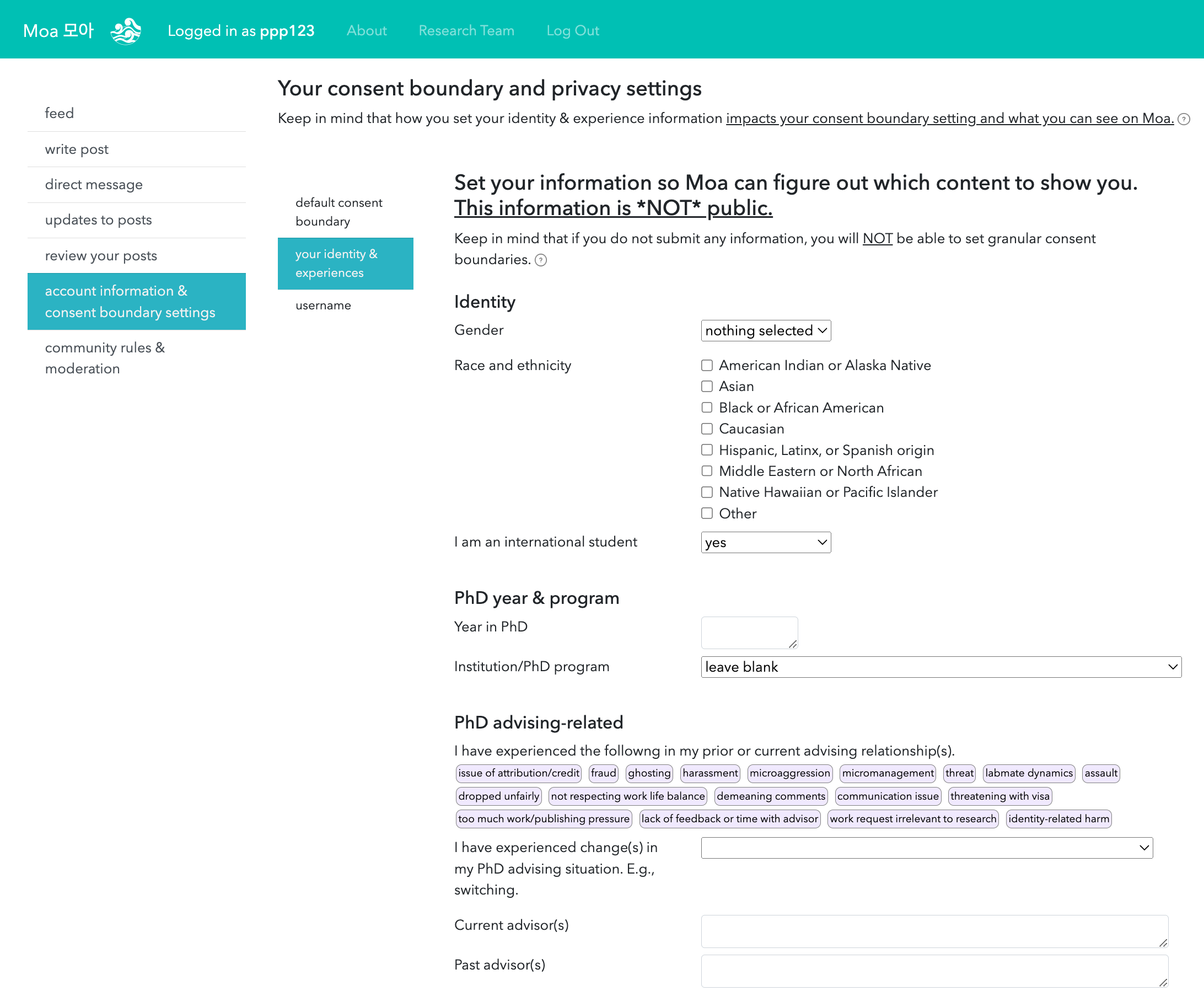}
    \caption{Users can update their identity traits and experiences on Moa in the account settings page.}
    \label{fig:moa-declare-traits}
\end{figure*}

\begin{figure*}[t!]
    \centering
    \includegraphics[width=.7\linewidth]{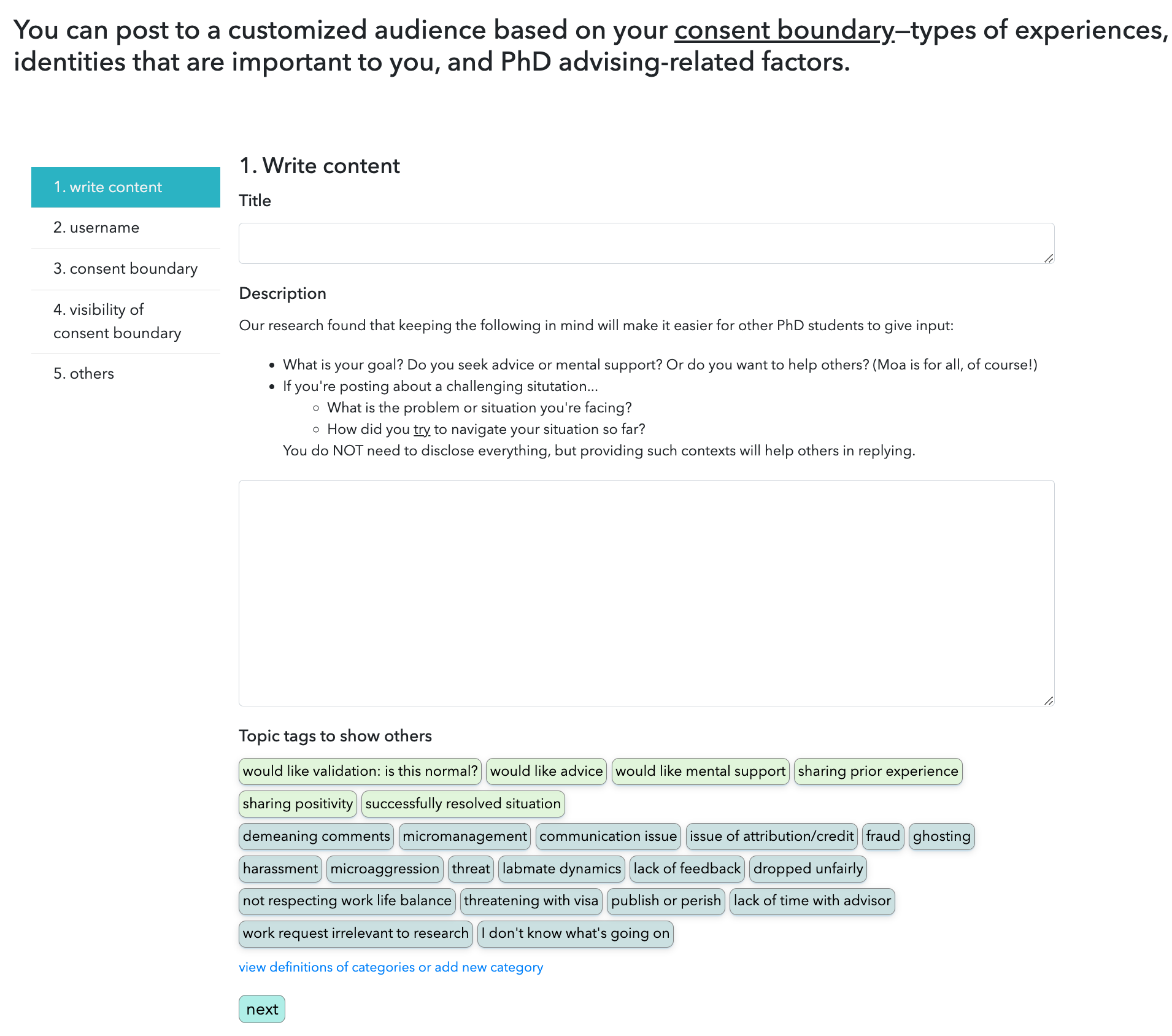}
    
    \vspace{10px}
    
     \includegraphics[width=.65\linewidth]{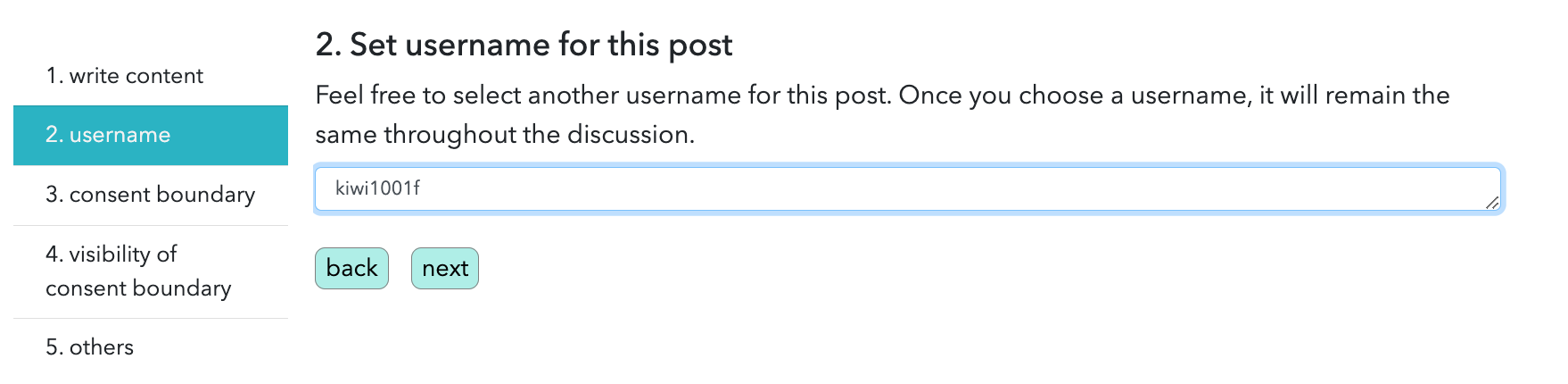}
    \caption{Moa's posting interface. Users are prompted to think about the goal of why they are posting, and can set a different username. The rest of the posting interface is shown in Figure \ref{fig:post2}.}
    \label{fig:post-1}
\end{figure*}

\begin{figure*}[t!]
    \centering
     \includegraphics[width=.7\linewidth]{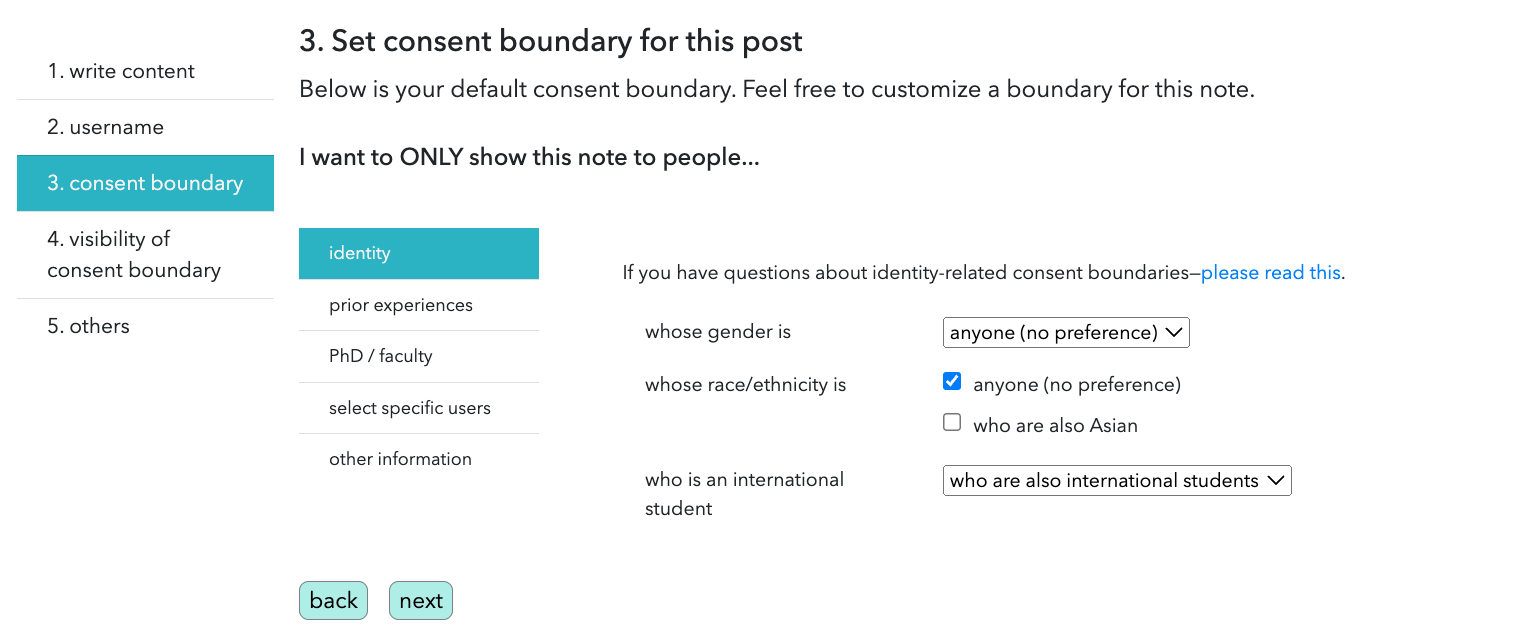}
     \includegraphics[width=.7\linewidth]{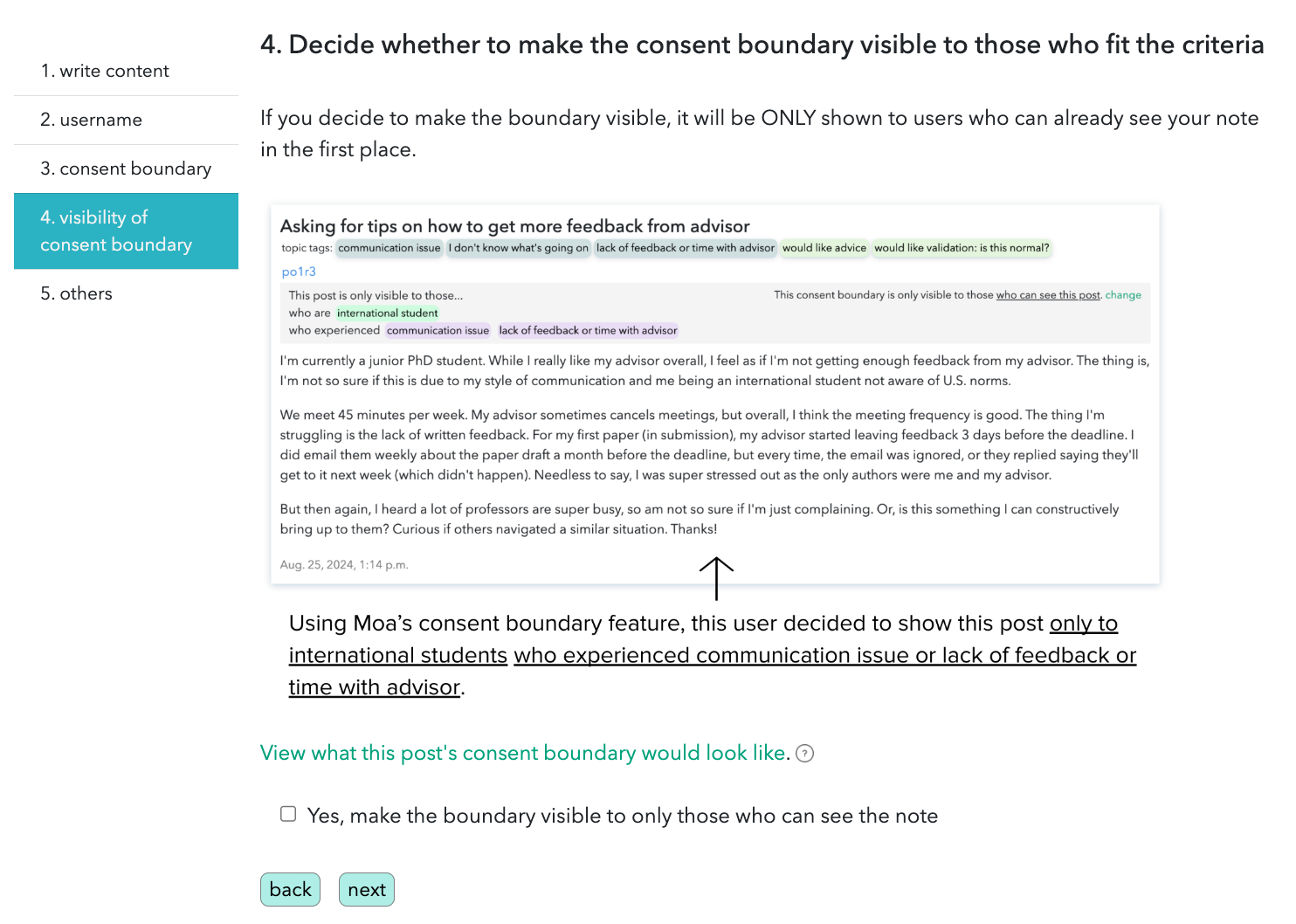}
    \caption{Moa's posting interface (continued from Figure \ref{fig:post-1}). A user can set a specific consent boundary per post and also decide whether to show the boundary to those who can see the post.}
     \label{fig:post2}
\end{figure*}

\end{document}